\documentclass[pdflatex,sn-mathphys-num]{sn-jnl}% Math and Physical Sciences Numbered Reference Style
\usepackage{graphicx}%
\usepackage{multirow}%
\usepackage{amsmath,amssymb,amsfonts}%
\usepackage{amsthm}%
\usepackage{mathrsfs}%
\usepackage[title]{appendix}%
\usepackage{xcolor}%
\usepackage{textcomp}%
\usepackage{manyfoot}%
\usepackage{booktabs}%
\usepackage{algorithm}%
\usepackage{algorithmicx}%
\usepackage{algpseudocode}%
\usepackage{listings}%
\theoremstyle{thmstyleone}%
\theoremstyle{thmstyletwo}%

\theoremstyle{thmstylethree}%

\begin{document}

\title[Article Title]{Minimal Proper Time and Deterministic Microstates: Emergent Quantum Fields and Relativistic Spacetime}

%%=============================================================%%
%% GivenName	-> \fnm{Joergen W.}
%% Particle	-> \spfx{van der} -> surname prefix
%% FamilyName	-> \sur{Ploeg}
%% Suffix	-> \sfx{IV}
%% \author*[1,2]{\fnm{Joergen W.} \spfx{van der} \sur{Ploeg} 
%%  \sfx{IV}}\email{iauthor@gmail.com}
%%=============================================================%%

\author*[1,2]{\fnm{Alessio Maiezza} \sur{}}\email{alessiomaiezza@gmail.com}

\affil*[1]{Dipartimento di Scienze Fisiche e Chimiche, Universit\`a degli Studi dell'Aquila, via Vetoio, I-67100, L'Aquila, Italy}

\affil[2]{INFN, Laboratori Nazionali del Gran Sasso, 67010 Assergi, L'Aquila, Italy}

%%==================================%%
%% Sample for unstructured abstract %%
%%==================================%%

\abstract{We develop a top-down counterpart of the minimal proper-time formulation of quantum field theory previously introduced as an effective bottom-up framework. Starting from a deterministic pre-geometric substrate of causally ordered events, we show how coarse-graining over microscopic histories leads, at low energies, to an effective Nambu-like quantum dynamics. The elementary deterministic update is identified with the minimal proper-time step, while the growth of coarse-grained equivalence classes controls both the ultraviolet dissipative correction and the scale dependence of the effective quantization strength, encoded in a running Planck constant. In this way, the proper-time cutoff kernel of the bottom-up formulation acquires a microscopic interpretation as the inverse growth of unresolved deterministic histories. In the infrared limit, the dissipative term vanishes and standard unitary quantum field theory is recovered. The same coarse-grained structure also provides a natural setting for an emergent relativistic spacetime geometry, compatible in the macroscopic limit with Einstein gravity. The resulting picture suggests a common deterministic origin for minimal-scale structure, quantum behavior, and relativistic spacetime.}


%%\pacs[JEL Classification]{D8, H51}

%%\pacs[MSC Classification]{35A01, 65L10, 65L12, 65L20, 65L70}

\maketitle

\section{Introduction}

A long-standing tension between quantum theory and relativistic covariance is the distinguished role played by time in the usual Hamiltonian formulation. In ordinary quantum mechanics (QM), and in the Schr\"odinger representation of quantum field theory (QFT), time appears as an external evolution parameter, rather than as a coordinate on the same footing as the spatial ones. An elegant way to soften this distinction was proposed by Feynman~\cite{PhysRev.76.749} and Nambu~\cite{Nambu:1950rs}, who showed that the usual time evolution can be recovered from a formulation based on an invariant proper-time parameter. In this approach, ordinary Schr\"odinger evolution is not taken as fundamental, but emerges after imposing the proper-time constraint. This perspective is particularly useful because it provides a natural language in which relativistic covariance and microscopic evolution can be discussed simultaneously.

A related idea, which has often appeared in the literature, is the existence of a fundamental scale of Nature. The typical realization of this idea is a minimal length associated with the Planck scale~\cite{Mead:1964zz,GARAY_1995,Kempf_1995,Padmanabhan:1996ap,Modesto_2009,Nicolini_2011,Bosso_2023,Bosso_2024}. In a relativistic setting, however, it is natural to formulate such a scale in terms of an invariant proper-time interval.

Recently, these ideas were combined in a QFT framework~\cite{Maiezza:2026wrp}, which will provide the starting point of the present work. That construction uses QFT in the Schr\"odinger representation~\cite{Hatfield:1992rz}, in order to keep the formalism close to the language of ordinary quantum mechanics. It first generalizes Nambu's proper-time formulation from quantum mechanics to QFT, and then introduces a fundamental scale through a minimal proper time within the Nambu-like framework.

The resulting theory goes beyond a QFT endowed with an intrinsic cutoff. One of its central features is the direct link between the minimal proper time and a controlled violation of unitarity at the ultraviolet scale. This effect is accompanied by a running effective Planck constant, suggesting that the strength of quantum behavior itself may become scale dependent. In particular, the high-energy regime can approach a deterministic limit, while standard quantum behavior is recovered at lower energies.

The aim of the present work is to propose a microscopic realization of this picture. We start from a deterministic pre-geometric substrate of causally ordered events and study the macroscopic dynamics obtained after coarse-graining over microscopic histories. In this sense, the present work may be viewed as the top-down counterpart of Ref.~\cite{Maiezza:2026wrp}, which provided the effective bottom-up formulation.

The approach developed here is inspired by cellular automata \textit{\`a la} 't Hooft~\cite{thooft2016cellular}. The minimal proper time naturally fits with a single discrete update of an automaton-like microscopic dynamics. At the microscopic level the evolution is deterministic and information-preserving. However, once one restricts the description to coarse-grained equivalence classes of indistinguishable configurations, the induced macroscopic evolution need not be exactly unitary. The apparent loss of unitarity is then not fundamental, but arises from the loss of microscopic distinguishability associated with coarse-graining.

This observation provides the key mechanism of the paper. The leakage between coarse-grained equivalence classes generates an effective dissipative contribution, controlled by the entropy variation of the corresponding microscopic sectors. If the number of unresolved microscopic histories grows as a power of the proper-time scale, the dissipative term is suppressed in the infrared and standard unitary QFT is recovered. Within the same construction, the effective Planck constant is determined by the unresolved fraction of microscopic histories. This leads to a microscopic interpretation of the proper-time cutoff kernel introduced in Ref.~\cite{Maiezza:2026wrp}: the kernel is identified with the inverse growth of the coarse-grained equivalence classes.

A further question is whether the same deterministic substrate can support an effective relativistic spacetime description. Since the fundamental events are not assumed to live on a pre-existing smooth manifold, the metric must also be reconstructed internally from the coarse-grained structure. We show that, in the macroscopic limit, the weighted covariance of coordinate fluctuations between interwoven event chains defines effective metric data. The corresponding smooth geometry is meaningful only in the infrared regime, where the dissipative corrections are suppressed. Under the standard infrared assumptions of locality, diffeomorphism covariance, second-order metric equations, and no additional long-range tensorial structures beyond the emergent metric, Lovelock's theorem then selects Einstein gravity as the effective macroscopic gravitational theory.

In this framework, the gravitational sector is therefore not introduced through a fundamental Einstein-Hilbert action. Rather, the metric is reconstructed from the statistical structure of the deterministic substrate, while the source tensor is defined as the response of the effective Nambu energy to local deformations of the emergent metric. The resulting gravitational equations should be understood as infrared effective equations, valid only after the smooth geometric description has emerged.

The idea that the strength of quantum behavior may itself be an effective, scale-dependent notion has also appeared in different forms in the literature.
For example, scenarios have been proposed in which the canonical commutator is associated with a symmetry that is broken only in the low-temperature phase,
so that Planck's constant effectively vanishes at high temperature~\cite{Hossenfelder:2012uy}.
Related ultraviolet classicalizing limits have also been discussed in the context of generalized uncertainty principles, where the canonical commutator or the
uncertainty relation is deformed~\cite{Ong:2018zqn}. Another complementary viewpoint is that $\hbar$ may be regarded as an effective parameter
characterizing the emergent Minkowski vacuum~\cite{Volovik:2009xs}, while in Adler's trace-dynamics program quantum theory itself arises as an emergent
statistical phenomenon from a deeper matrix dynamics~\cite{adler2004quantum}. The mechanism proposed here is different from these approaches: the running
effective Planck constant is not postulated through a thermal order parameter, a phenomenological deformation of the uncertainty principle, or a vacuum
parametrization, but is fixed by the unresolved fraction of microscopic deterministic histories inside a coarse-grained equivalence class.

In a broad sense, the present work also belongs to the line of thought associated with superdeterministic approaches to Nature. Even after Bell's seminal work~\cite{bell1964einstein}, superdeterminism has not been ruled out~\cite{brans1988bell} and has received renewed attention in the last two decades; see, for example, Refs.~\cite{tHooft:2001qty,Palmer:2008jh,Hall:2010zzf,Hossenfelder:2019shy,Donadi:2020aqz,tHooft:2020qfg,tHooft:2020tuu,Powers:2021rfg,Hance:2022juc,Palmer:2023vfw,Arroyo:2024saq}, in a non-comprehensive list. The possibility that Nature may approach a classical or deterministic regime at the Planck scale is also conceptually related to classicalization, although the logic followed here is different~\cite{Dvali:2011aa}; see also Ref.~\cite{DeAngelis:2022qhm} for another perspective.

The article is organized as follows. In Sec.~\ref{sec:resume}, we summarize the ingredients of Ref.~\cite{Maiezza:2026wrp} that are needed for the present construction. In Sec.~\ref{sec:setup}, we introduce the deterministic substrate of fundamental events, the microscopic update map, and the coarse-grained equivalence classes. In Sec.~\ref{sec:QFT-unit}, we show how an effective Nambu-like QFT and infrared unitary dynamics emerge from the coarse-grained description. In Sec.~\ref{subsec:heff}, we identify the effective Planck constant and the corresponding proper-time cutoff kernel. In Sec.~\ref{sec:GR}, we discuss the emergence of an effective relativistic geometry and its compatibility with Einstein gravity in the infrared limit. We offer some discussions in Sec.~\ref{sec:limitations} and conclude in Sec.~\ref{sec:end}. Two illustrative toy models are presented in Appendices~\ref{appendix-toy} and~\ref{appendix-gravity-toy}.

\section{The minimal proper time}\label{sec:resume}

In this section, we summarize the ingredients of \cite{Maiezza:2026wrp} that are needed for the top-down construction developed below. The key
point is the introduction of a minimal proper time, $\tau_{\min}$, in the Schr\"odinger representation of QFT. This scale softly relaxes the
Nambu-like constraint, leading to a controlled violation of unitarity at the ultraviolet scale and to a running effective Planck constant. In the
following sections, these effective quantities will be reinterpreted as emergent consequences of coarse-graining over deterministic microscopic
histories.

The functional formalism allows one to describe QFT in the Schr\"odinger representation, through a functional Schr\"odinger equation and, therefore,
to keep close contact with the language of ordinary quantum mechanics. This is useful for two reasons. First, it makes it possible to generalize
Nambu's original constraint from quantum mechanics to QFT \cite{Nambu:1950rs}. Second, once such a constraint is available, it can be softly relaxed by introducing a small but finite $\tau_{\min}$, whose effects can then be systematically studied. We identify $\tau_{\min}$ with the Planck time, $\sqrt{\hbar G/c^5}$.

One of the central results of Ref.~\cite{Maiezza:2026wrp} is the emergence of a running effective Planck constant. This suggests a possible scenario
in which quantum behavior is suppressed near the Planck scale, while the standard quantum regime is recovered at lower energies. Let us now
summarize the main technical points.

\subsection{QFT in Schr\"odinger representation}

To keep close contact with the language of the ordinary QM, we use QFT in the Schr\"odinger representation.

In QFT, the notion of coordinate $x$ is replaced with that of the field:
\begin{equation}
\hat{\phi}(\vec{x}) \lvert \phi \rangle = \phi(\vec{x}) \, \lvert \phi \rangle \,,
\end{equation}
where the hatted symbol indicates the operator, while the plain symbol denotes its eigenvalues, which is a function. 

The QM wave-function $\psi(\vec{x},t)$ is replaced by the wave-functional,
\begin{equation}
\Psi[\phi, t] \,,
\end{equation}
satisfying the  functional Schr\"odinger equation:
\begin{equation}\label{Seq1}
H \Psi[\phi, t] = i \frac{\partial}{\partial t} \Psi[\phi, t] \,.
\end{equation}
Leveraging the Dirac delta representation,
\begin{equation}
\frac{\delta \phi(\vec{x})}{\delta \phi(\vec{y})} = \delta(\vec{x}-\vec{y}) = \left[\frac{\delta}{\delta \phi(\vec{y})}, \phi(\vec{y}) \right]  \,,
\end{equation}
where $\frac{\delta}{\delta \phi(\vec{x})}$ denotes the functional derivative, and comparing the last commutator to the one in the canonical commutation relations (CCR) yields
\begin{equation}
\pi=- i \hbar \frac{\delta}{\delta \phi(\vec{x})} \,,
\end{equation}
where $\pi$ is the conjugate field. 

Then, the Hamiltonian $H$ for a real scalar free field reads,
\begin{equation}\label{Seq2}
H = \int d^3x \left[ -\frac{\hbar^2}{2} \frac{\delta^2}{\delta \phi(\vec{x})^2} + \frac{1}{2} \left( \nabla \phi(\vec{x})^2  + m^2 \phi(\vec{x})^2 \right) \right] \,,
\end{equation}
defining \eqref{Seq1}.

\subsection{Nambu's constraint and its controlled violation}

The minimal setup above provides the ingredients to mimic Nambu's approach \cite{Nambu:1950rs} now in QFT.

Using \eqref{Seq1} and \eqref{Seq2}, one can propose the equation,
\begin{equation}\label{eq:nambu_qft}
H' \Psi[\phi, t,\tau]= i \frac{\partial}{\partial \tau} \Psi[\phi, t, \tau] \,,
\end{equation}
with
\begin{equation}
H'=H-i\frac{\partial}{\partial t} \,,
\end{equation}
being $\tau$ the proper time.  For recent approaches to renormalization group in terms of proper time, see \cite{Abel:2023ieo,Bonanno:2025tfj,Bonanno:2025qsc,Giacometti:2025qyy}.

We shall take the Planck constant $\hbar=1$,  except when an explicit reintroduction becomes helpful.

As usual, it is convenient to transform \eqref{eq:nambu_qft} into an eigenvalues problem by assuming the solution of the form:
\begin{equation}
\Psi[\phi, t,\tau]= \Psi_{\lambda}[\phi, t] e^{-i \lambda \tau} \,,
\end{equation}
leading to,
\begin{equation}
H' \Psi_{\lambda}[\phi, t] = \lambda \Psi_{\lambda}[\phi, t]\,.
\end{equation}
The crux of the Nambu approach is to obtain the usual Schr\"odinger equation by integrating over $\tau$ the wave-functional $\Psi_{\lambda}[\phi, t] e^{-i \lambda \tau}$, which leads to
\begin{equation}
\Psi^R[\phi, t] = \text{Re}  \left(\int_{0}^{\infty} d\tau\, \Psi[\phi, t,\tau]\right)\ = \pi \delta(\lambda)\, \Psi_{\lambda}[\phi, t] \,.
\end{equation}
It is easy to show that the physical states satisfy \cite{Nambu:1950rs,Maiezza:2026wrp}:
\begin{equation}\label{constraint}
H' \Psi^R[\phi, t] = 0  \hspace{2em} \Leftrightarrow \hspace{2em}   \lambda=0 \,.
\end{equation}
where the superscript $R$ denotes the 'Real' world. 

This reformulation treats $t$ as a  label rather than a fundamental parameter of evolution: this emerges only after the constraint. 
In this way, the formulation is naturally compatible with the diffeomorphism-invariant spirit of GR.

\subsection{The impact of minimal proper time}

The introduction of a non-zero but small minimal proper time, $\tau_{\min}>0$, implies a modification of the integration over $\tau$. For example,
the Nambu-like constraint is softly relaxed as,
\begin{equation}\label{approx_viol}
\text{Re}\left[\int_{\tau_{\min}}^\infty e^{-i\lambda\tau} d\tau\right] = \pi\delta(\lambda) - \tau_{\min} + \mathcal{O}(\tau_{\min}^3) \,,
\end{equation}
valid for physically small $\tau_{\min}$. The expression in \eqref{approx_viol} also enables non-zero $\lambda$ as the 'Real' world wave-functional:
\begin{align}\label{superposition}
& \Psi[\phi,t,\tau]=\int_{-\infty}^\infty d\lambda\, f(\lambda)\, \Psi_\lambda[\phi,t]\, e^{-i \lambda \tau}   \nonumber \\
&\Psi_{R}[\phi,t] = \int_{\tau_{\min}}^\infty d\tau \int_{-\infty}^\infty d\lambda\, f(\lambda)\, \Psi_\lambda[\phi,t]\, e^{-i \lambda \tau} \,,
\end{align}
with
\begin{equation}
f(\lambda) = \int \mathcal{D}\phi \, \Psi[\phi,0,0] \, \Psi_\lambda^*[\phi,0] \,.
\end{equation}
As it must be, if it were $\tau_{\min}\rightarrow 0$, one would have the exact constraint, $\Psi_{(\lambda=0)}$; vice versa, if it were $f(\lambda)=\delta(\lambda)$, one would obtain back $\Psi^R=\Psi_{(\lambda=0)}$ and $\tau_{\min}=0$ \footnote{Here we understand that the $\tau_{\min}$ dependence in $ f(\lambda)$ originates from a necessary regularization of the functional -- see \cite{Maiezza:2026wrp} for the details.}. 

For the present purposes, the relevant result can be written as,
\begin{equation}\label{gen3}
\Psi_{R}[\phi,t] = \Psi_{\lambda=0}[\phi,t] \left[ 1- a\,\tau_{\min} \tilde{f}(t) \right]+ \mathcal{O}(\tau_{\min}^2) \,,
\end{equation}
with
\begin{equation}
\tilde{f}(t) := \int_{-\infty}^{\infty} d\lambda f(\lambda) e^{i \lambda t} \,,
\end{equation}
and $a$ is dimensionful, arbitrary constant. The arbitrariness of $a$ reflects the freedom in normalizing the Nambu-like  constraint: a rescaling of the $\delta(\lambda)$ distribution in \eqref{constraint} leaves the physical projection onto $\lambda=0$ unchanged. This normalization 
freedom is not an ambiguity of the formulation but a structural feature,  analogous to the choice of normalization for eigenstates in quantum mechanics.  As we shall show in Sec.~\ref{subsec:heff}, it is fixed by the microscopic top-down matching condition.

Importantly, \eqref{gen3} implies a controlled, suppressed violation of unitarity at the Planck scale:
\begin{equation}
U \mapsto U \left( 1- a\,\tau_{\min} \tilde{f}(t) \right) + \mathcal{O}(\tau_{\min}^2) \,,
\end{equation}
which, in turn, entails an effective Hamiltonian, $H_{eff}= H-i D$ (where $D$ denotes the dissipative part) and a modification of CCR:
\begin{equation}\label{deforme}
[\hat{\phi}(\vec{x}), \hat{\pi}(\vec{y})] = i\hbar\, \delta(\vec{x} - \vec{y}) \left(1 - 4 a\,\tau_{\min}  \tilde{f}(t)\right) + \mathcal{O}(\tau_{\min} ^2) \,.
\end{equation}
This enables one to define a running effective Planck constant:
\begin{equation}\label{hrun2}
\hbar_{\rm eff}(t) := \hbar \left(1 - 4 a\, \tau_{\min} \tilde{f}(t)\right) + \mathcal{O}(\tau_{\min}^2) \,.
\end{equation}
This result and its interpretation are the conceptual core of Ref.~\cite{Maiezza:2026wrp}. At this bottom-up level, \eqref{hrun2} should be understood as an effective parametrization: the normalization $a$ reflects the freedom in the normalization of the
Nambu-like constraint, while the profile $\tilde f(t)$ encodes the strength of the off-constraint contribution. The possibility that
$\hbar_{\rm eff}$ becomes strongly suppressed at very short times, or equivalently at energies of order $1/t\simeq M_{\rm Pl}$, is therefore
an input of the effective construction. Therefore, the intriguing possibility of a fundamentally deterministic universe motivates the search for a theory capable of implementing this feature from first principles. This is the object of the rest of this work.

It is worth emphasizing that similar ideas, in which quantum behavior becomes scale-dependent or emergent, have been discussed in Refs.~\cite{Hossenfelder:2012uy,
Ong:2018zqn,Volovik:2009xs,adler2004quantum}. The construction developed below is different in that the running of \(\hbar_{\rm eff}\) will be derived
from the unresolved fraction of deterministic microscopic histories inside a coarse-grained equivalence class. In particular, the matching in
Subsec.~\ref{subsec:heff} will fix both $a$ and $\tilde f$ in terms of the coarse-grained cardinality.

\section{Fundamental Events, Microstates, and Coarse-Graining}\label{sec:setup}

The central idea is that the most fundamental objects in Nature are not spacetime points, but correlations among elementary events. What is
observed at macroscopic scales is therefore a coarse-grained pattern of relations among such events. We shall postulate that these microscopic
correlations are fundamentally deterministic, while QFT emerges statistically from the coarse-grained description of the deterministic
substrate.

Let us consider an ordered set of fundamental events:
\begin{equation}\label{event_order}
\mathcal{E}^c = \{ e_0^c, e_1^c, e_2^c, \dots \}, \quad e_i^c \prec e_j^c \text{ for } i<j \,,
\end{equation}
where the superscript $c$ denotes one among an arbitrarily large number of different chains. We shall suppress this superscript, for simplicity,
except when explicitly required. These events are elements of a fundamental substrate (\textit{a priori}, not yet space or spacetime). Each event $e_n$ carries a fundamental microscopic state $\phi(e_n) \in \Lambda$.

\subsection{Deterministic Dynamics and Microscopic Hilbert Space}

At the fundamental level, we assume a deterministic, bijective map $F:\Lambda \to \Lambda$ that updates the microstates from one event to the next:
\begin{equation}
\phi(e_{n+1}) = F(\phi(e_n)) \,.
\end{equation}
We identify this elementary discrete step $n \to n+1$ with the minimal proper-time interval $\tau_{\min}$.

Following \cite{thooft2016cellular}, we represent this deterministic dynamics in a Hilbert-space language by associating each microscopic
configuration with an orthonormal basis vector, $\phi(e_n) \to \vert \phi_n \rangle$. At this stage, the Hilbert space should be understood as a linear representation of the deterministic state space, rather than as the origin of quantum behavior. The
deterministic evolution is then implemented by a linear operator $\hat{F}$ such that:
\begin{equation}\label{cyclic}
\hat{F} \vert \phi_n \rangle = \vert \phi_{n+1} \rangle = \vert F(\phi_n) \rangle \,. 
\end{equation}
This constitutes a formal linearization of the underlying deterministic system.

The fundamental transition amplitude for a single step is purely deterministic and is given by the matrix elements of $\hat{F}$:
\begin{equation}\label{Adet}
\langle \phi'\vert \hat{F} \vert \phi \rangle = \delta_{\phi',F(\phi)} \,,
\end{equation}
where $\delta$ is the Kronecker delta. For an evolution over $N$ discrete steps, corresponding to a total proper time $\tau = N \tau_{\min}$, the
exact microscopic amplitude is determined by the $N$-th power of the evolution operator:
\begin{equation}\label{AdetN}
\langle \phi'\vert \hat{F}^N \vert \phi \rangle = \delta_{\phi',F^{(N)}(\phi)} \,,
\end{equation}
where $F^{(N)} = F \circ \cdots \circ F$ denotes the map iterated $N$ times.

It is crucial to emphasize that the deterministic map $F$ is not restricted to act along a single isolated chain. Restoring the chain index $c$ introduced in Eq.~\eqref{event_order}, the map can connect microstates belonging to distinct chains, namely
\begin{equation}\label{interweaving}
F(\phi(e_n^c)) = \phi(e_{n+1}^{c'}) \,, \quad \text{with } c \neq c' \,.
\end{equation}
This topological interweaving among different chains prevents the substrate from reducing to a trivial collection of non-communicating one-dimensional histories. As we shall see, the mixing induced by $F$ is a prerequisite for the emergence of an effectively multidimensional macroscopic geometry and of the metric structure required for a realistic QFT.

\subsection{Coarse-Graining and Classical Averaging}

Let us now introduce the statistical effects associated with the loss of microscopic distinguishability at lower energies.

Specifically, we define the dimensionless coarse-graining scale
$$
\mu=\frac{\tau-\tau_{\min}}{\tau_{\min}}\,,
$$
with $\tau=N\tau_{\min}$ and $N$ a positive integer. We then introduce a coarse-graining map $\mathcal{C}_\mu$ acting on microscopic configurations,
\begin{equation}
\phi_\mu := \mathcal{C}_\mu(\phi) \,.
\end{equation}
The microscopic state $\phi$ is identified with $\phi_0$, namely the configuration resolved at $\mu=0$. The corresponding state in the
Hilbert-space representation will be denoted by $\vert \phi_\mu \rangle$.

We define the set of microscopic configurations that are indistinguishable at resolution $\mu$, namely the equivalence class,
\begin{equation}
[\phi_\mu] = \{ \phi \in \Lambda \mid \mathcal{C}_\mu(\phi) = \phi_\mu \}.
\end{equation}
Its cardinality is
\begin{equation}
n_\mu := |[\phi_\mu]| \in \mathbb{N} \,. 
\end{equation}
As $\mu$ increases, the resolution decreases and the number of microscopic configurations associated with the same macroscopic state
grows. Physically, as the number of elementary evolution steps $N$ increases, the recurrent interweaving induced by $F$ progressively mixes
the exact microscopic information. As a consequence, an increasingly large number of deterministic histories becomes indistinguishable to a
macroscopic observer, causing the cardinality $n_\mu$ to grow with $\mu$.

In the Hilbert-space representation, the coarse-grained state is described by the normalized average over the microscopic configurations belonging
to the same equivalence class:
\begin{equation}\label{fi-med}
\vert \phi_\mu \rangle = \frac{1}{\sqrt{n_\mu}}  \sum_{\phi \in [\phi_\mu]} \vert \phi \rangle \,.
\end{equation}
This expression should be understood as the linear representation of a macroscopic equivalence class, not as a fundamental quantum superposition
of microscopic states. Notice that here we are omitting an index, for brevity, since the explicit notation would be
$\mathcal{C}_\mu \vert \phi_n \rangle = \vert \phi_{\mu,n} \rangle$. We shall continue to omit this index $n$, except in cases where explicit
notation is necessary.

\section{Emergence of QFT and Effective Unitary Dynamics}\label{sec:QFT-unit}

At the microscopic level, the deterministic evolution strictly preserves the purity of the states. The map $\hat F$ is a linear representation of a bijection on the microscopic configurations, and therefore no information is lost at the fundamental level. However, once the description is restricted to coarse-grained equivalence classes of indistinguishable configurations, the induced macroscopic evolution need not be isometric. In this sense, the apparent loss of unitarity is an effective property of the coarse-grained description, rather than a fundamental violation of the deterministic dynamics.

We now evaluate the emergent dynamics on the coarse-grained Hilbert space. The transition amplitude between two macroscopic states over a single fundamental time step $\tau_{\min}$ is obtained by sandwiching the exact microscopic operator $\hat F$ between the macroscopic states defined in Eq.~\eqref{fi-med}:
\begin{equation}\label{K1}
K(\mu \to \mu'; \tau_{\min}) =
\langle \phi_{\mu'} \vert \hat{F} \vert \phi_\mu \rangle
= \frac{1}{\sqrt{n_\mu n_{\mu'}}}
\sum_{\phi \in [\phi_\mu]}
\sum_{\phi' \in [\phi_{\mu'}]}
\delta_{\phi',F(\phi)} \, .
\end{equation}
Since $F$ interweaves different chains, $c\neq c'$, the fundamental evolution mixes microscopic configurations belonging to distinct causal histories. Consequently, a macrostate $\vert \phi_\mu\rangle$ does not necessarily map into a single coarse-grained macrostate, but can have support on several equivalence classes.

Moreover, even for the dominant causal transitions, the connected equivalence classes generally possess different cardinalities, $n_\mu\neq n_{\mu'}$. This structural mismatch, together with the transversal mixing induced by $F$, implies that the map induced on the coarse-grained subspace need not be an isometry. The corresponding macroscopic evolution is therefore effectively non-unitary, even though the underlying microscopic dynamics remains deterministic and information-preserving. Strict unitarity is recovered in the regime in which the relative leakage between equivalence classes becomes negligible, namely when $\Delta n_\mu/n_\mu\to 0$; in the infrared this is realized as $n_\mu\to\infty$.

The effective action of $\hat F$ on the macroscopic subspace can then be parameterized in terms of an emergent Hermitian generator together with a non-Hermitian dissipative contribution. This is analogous to the effective description of open quantum systems and establishes a direct contact with the framework of Ref.~\cite{Maiezza:2026wrp}, summarized in Sec.~\ref{sec:resume}. We take the dissipative contribution to be governed by the discrete change in the Boltzmann entropy of the coarse-grained equivalence classes,
\begin{equation}\label{Bol}
S_\mu = \ln(n_\mu) \, .
\end{equation}
Over a single fundamental step $\tau_{\min}$, let $\Delta n_\mu$ denote the number of microscopic configurations whose image under $F$ leaves the equivalence class $[\phi_\mu]$ and falls into adjacent coarse-grained classes. When this number is small compared with the total cardinality of the class, $\Delta n_\mu\ll n_\mu$, the corresponding entropy variation is
\begin{equation}\label{DeltaBol}
\Delta S_\mu
=
\ln\left(1+\frac{\Delta n_\mu}{n_\mu}\right)
\approx
\frac{\Delta n_\mu}{n_\mu} \, .
\end{equation}
Here $n_\mu$ counts the total number of microscopic configurations in the coarse-grained class $[\phi_\mu]$, while $\Delta n_\mu$ counts only the boundary configurations whose image under the deterministic map crosses from $[\phi_\mu]$ to neighboring equivalence classes. Thus $n_\mu$ is a bulk quantity, whereas $\Delta n_\mu$ is a boundary quantity.

For the class of local bijective, permutation-like maps considered here, the number of such boundary-crossing configurations per elementary step remains bounded as the bulk cardinality grows. We therefore assume
\begin{equation}\label{bounded-leakage}
\Delta n_\mu = \mathcal{O}(1) \, ,
\end{equation}
independently of the size of the coarse-grained class. This assumption is not a consequence of bijectivity alone, but of the locality of the map with respect to the coarse-graining structure. The cyclic permutation model of Appendix~\ref{appendix-toy} provides an explicit realization of this situation. This boundary behavior should be distinguished from the growth of the bulk cardinality of the coarse-grained class, which will scale as $n_\mu\sim N^p$ at large proper time.

The coarse-grained transition amplitude over a single fundamental step can then be parameterized as
\begin{equation}\label{Evolution}
\langle \phi_{\mu'} \vert \hat{F} \vert \phi_\mu \rangle
\approx
\langle \phi_{\mu'} \vert
\exp\left[-i\tau_{\min}H' - \Delta S_\mu\right]
\vert \phi_\mu\rangle
\approx
\langle \phi_{\mu'} \vert
\exp\left[-i\tau_{\min}H' - \frac{\Delta n_\mu}{n_\mu}\right]
\vert \phi_\mu\rangle \, ,
\end{equation}
where
\begin{equation}
H' = H - i\frac{\partial}{\partial t}
\end{equation}
is the extended generator driving the dynamics in proper time. In this effective description, the non-unitary part of the coarse-grained evolution is encoded in the amplitude-suppression factor $e^{-\Delta S_\mu}$. The entropy variation therefore measures the loss of microscopic distinguishability induced by coarse-graining and controls the dissipative correction. Appendix~\ref{appendix-toy} provides a simple toy model in which this construction can be explicitly realized.

This dissipative correction determines the controlled violation of unitarity in the coarse-grained dynamics and naturally vanishes as the proper time flows toward macroscopic scales. In the proper-time formalism, physical amplitudes are obtained by integrating over all possible proper-time durations $\tau$, corresponding to a discrete number of fundamental steps
\begin{equation}
N = \frac{\tau}{\tau_{\min}} \, .
\end{equation}
We assume a power-law scaling for the cardinality of the coarse-grained equivalence class resolved at the final proper-time scale,
\footnote{%
The power-law growth may be motivated by the causal structure of the microscopic substrate. After $N$ elementary steps, the map $F$ can interweave chains that lie within a causal neighborhood of radius $\mathcal{O}(N)$ in the discrete event graph. The number of microscopic configurations within such a neighborhood scales as a $d$-dimensional volume, $n_\mu\sim N^d$, where $d$ is the effective topological dimension of the network. The exponent $p$ is therefore a measure of the effective dimensionality of the microscopic causal structure.}
\begin{equation}\label{n-growth}
n_\mu(N)\simeq N^p \, .
\end{equation}
We stress that the coarse-graining is not applied after each elementary step. The reason is that the fundamental evolution is assumed to be the exact deterministic evolution generated by $\hat F$. The coarse-grained description is instead a macroscopic readout performed only after the microscopic history has evolved for \(N\) elementary steps. Thus one first computes the exact deterministic iterate \(\hat F^N\), and only then projects the result onto the final equivalence class resolved at the scale $\mu(N)$. In this sense, $n_\mu(N)$ denotes the cardinality
of the final coarse-grained class, not a step-dependent variable updated during the microscopic evolution.

This prescription is important for obtaining the correct infrared limit. If the coarse-graining were applied after every elementary step,
the projected evolution would define an autonomous effective dynamics on the coarse-grained classes. Such a dynamics would be Markovian, or
open-system-like, at the macroscopic level: the microscopic correlations retained by the exact deterministic iterate \(\hat F^N\) would be erased at each step. The leakage factor would then be accumulated locally along the projected trajectory, rather than being estimated relative to the final coarse-grained class. In general this would leave a residual dissipative contribution in the infrared, unless an additional fine-tuning or renormalization of the coarse-grained transition rule were introduced.

The prescription used here avoids this problem. The microscopic dynamics remains deterministic and information-preserving throughout the whole $N$-step evolution, while the non-unitary contribution appears only as a final loss of microscopic distinguishability associated with the macroscopic projection. With $n_\mu(N)\sim N^p$ and bounded boundary leakage per elementary step, the cumulative leakage fraction scales as
$N/n_\mu(N)\sim N^{1-p}$. Therefore, for \(p>1\), the dissipative correction vanishes in the infrared and the standard unitaryproper-time evolution is recovered.

Using Eq.~\eqref{bounded-leakage}, the number of boundary-crossing configurations per elementary step is bounded, $\Delta n_\mu=\mathcal{O}(1)$. Hence the cumulative leakage fraction over $N$ microscopic steps scales as
\begin{equation}\label{cumulative-leakage}
N\frac{\Delta n_\mu}{n_\mu(N)}
\simeq
\frac{\gamma}{N^{p-1}} \, ,
\qquad
\gamma=\mathcal{O}(1) \, .
\end{equation}
The coefficient $\gamma$ depends on the microscopic boundary structure of the equivalence classes and can be absorbed into the normalization of the effective dissipative correction. Equivalently, one may set $\gamma=1$ when only the scaling behavior is relevant.

Evaluating the $N$-th power of the exact microscopic operator $\hat F$, one obtains
\begin{equation}\label{U-nonU}
\langle \phi_{\mu'} \vert \hat F^N \vert \phi_\mu\rangle
\approx
\left\langle \phi_{\mu'}\left\vert
\exp\left[-iN\tau_{\min}H'
-
N\frac{\Delta n_\mu}{n_\mu(N)}
\right]
\right\vert \phi_\mu\right\rangle
\approx
\left\langle \phi_{\mu'}\left\vert
\exp\left[-i\tau H'
-
\frac{\gamma}{N^{p-1}}
\right]
\right\vert \phi_\mu\right\rangle \, .
\end{equation}
In the deep ultraviolet regime, $N=\mathcal{O}(1)$ and $\tau=\mathcal{O}(\tau_{\min})$, the dissipative term is unsuppressed. It therefore suppresses short-distance contributions and acts as a dynamical ultraviolet regulator. Conversely, as the proper time flows toward the infrared limit, $N\to\infty$, the dissipative term vanishes provided
\begin{equation}\label{p-condition}
p>1 \, .
\end{equation}
Under this condition one obtains
\begin{equation}\label{IR-unitarity}
\lim_{N\to\infty}\hat F^N\big|_{\rm macro}
=
\exp(-iH'\tau)
:=
U(\tau) \, ,
\end{equation}
where $U(\tau)$ is the unitary evolution operator in proper time.\footnote{$H'$ must be made self-adjoint by proper boundary conditions \cite{Maiezza:2026wrp}.}

The condition $p>1$ is therefore essential: it guarantees that, for $\tau\gg\tau_{\min}$, the entropy-driven suppression decays fast enough to allow coherent propagation. If $p\leq 1$, the coarse-grained theory would remain dissipative even at macroscopic scales. Thus, standard unitary QFT dynamically emerges in the infrared limit of the proper-time integration, while the effective non-unitarity induced by coarse-graining is confined to the ultraviolet regime. See Appendix~\ref{appendix-toy} for an explicit simplified model.

\paragraph{Remarks.}

This brings us to an important consequence of the framework: the parameterized effective evolution depends on the macroscopic state
$\mu$ through the number of microstates $n_\mu$. In standard QFT in the Schr\"odinger representation, time evolution is governed by a universal, state-independent operator, which ensures the linear superposition principle. In contrast, the explicit $\mu$-dependence of the dissipative factor introduces an effective non-linearity into the coarse-grained dynamics.

However, the breakdown of the superposition principle at this level is neither problematic nor unexpected. It should also be distinguished from generic non-linear modifications of quantum mechanics, which are known to raise potential causality issues, including superluminal-signalling problems in the presence of entanglement~\cite{gisin1990weinberg}. In the present construction, the state-dependent non-linearity is not an independent modification of the Schr\"odinger equation acting on the infrared Hilbert space. Rather, it is induced by the dependence of the
coarse-grained leakage factor on the macroscopic equivalence class, and is therefore tied to the same quantity that controls the dissipative
correction. As shown above, this correction scales cumulatively as $\mathcal{O}(1/N^{p-1})$, and is unsuppressed only in the deep UV
regime $(N=\mathcal{O}(1) \,,\tau=\mathcal{O}(\tau_{\min}))$. At this fundamental scale, the continuous spacetime description underlying
standard QFT is no longer expected to be valid, and there is no reason to impose exact linear quantum mechanics as a fundamental principle.

This observation better clarifies why the non-linear effects do not accumulate into an infrared pathology. The non-linearity is controlled by
the same leakage fraction that suppresses the non-unitary part of the coarse-grained evolution. For a single elementary step one has
\begin{equation}
\frac{\Delta n_\mu}{n_\mu(N)} \sim N^{-p}\,,
\end{equation}
while over a proper-time interval $\tau=N\tau_{\min}$ the cumulative
effect scales as
\begin{equation}
N\,\frac{\Delta n_\mu}{n_\mu(N)} \sim N^{1-p}\,.
\end{equation}
Therefore, for $p>1$, both the dissipative correction and the state-dependent non-linear contribution are washed out in the infrared. The effective evolution then becomes state-independent, linear and unitary, so that the usual relativistic QFT causal structure is
recovered.

At intermediate scales, where $N$ is large but finite, residual non-linear corrections may in principle survive as suppressed asymptotic effects of order $N^{1-p}$. Such effects would represent possible signatures of the coarse-grained deterministic substrate rather
than inconsistencies of the infrared theory. They cannot accumulate into a finite IR violation within the present scaling regime, because their cumulative contribution decreases with $N$ whenever $p>1$. They may only become relevant near the fundamental proper-time scale, where the continuum QFT description itself is no longer assumed to be valid.

\subsection{Nambu-Schr\"odinger equation}

One can move to a basis that diagonalizes the effective macroscopic action of $\hat{F}^N$ in Eq.~\eqref{U-nonU}, namely from the macroscopic basis
$\vert \phi_{\mu,k}\rangle$ to $\vert \bar{\phi}_{\mu,k}\rangle$, such that
\begin{equation}
\hat{F}^N \vert \bar{\phi}_{\mu,k}\rangle
\approx e^{-i h_k \tau} \vert \bar{\phi}_{\mu,k}\rangle \,.
\end{equation}
Here the quantities $h_k$ are, in general, complex because the coarse-grained macroscopic evolution contains an effective dissipative component in the UV. As discussed above, this imaginary contribution vanishes in the continuum IR limit ($N \to \infty$), so that $e^{-i h_k \tau}$ becomes a pure phase. This behavior is also illustrated explicitly in the toy model of Appendix~\ref{appendix-toy}. This step shows how the underlying deterministic evolution can provide the complex phases required for quantum interference, in contrast with purely real diffusive dynamics\footnote{The pure phase part in $e^{-i h_k \tau}$ is guaranteed by the cyclic action of the deterministic operator $\hat{F}$ defined in Eq.~\eqref{cyclic} \cite{thooft2016cellular} (e.g., ``The Cogwheel Model''), together with the fact that the macroscopic amplitude is obtained by evaluating $\hat{F}$ on coarse-grained linear representations of equivalence classes, as illustrated in Appendix~\ref{appendix-toy}.}.

\medskip

The connection to the continuous field theory is established in the limit $\tau \gg \tau_{\min}$. The discrete event index $n$ is then mapped to the continuous proper-time parameter $\tau=n\,\tau_{\min}$. In this continuum description, the microscopic state $\phi(e_n^c)$ is represented as a field configuration over the coordinates $x=(t,\vec{x})$,
\begin{equation}
\phi(e_n^c) \rightarrow \phi(x,\tau) \equiv \phi(\vec{x},t,\tau) \,.
\end{equation}
Here the chain label $c$ is kept explicitly to recall that the emergent configuration arises from the collection of microscopic event chains.
Before the physical constraint is imposed, the effective description can therefore be viewed as living in an extended space of variables
$(x,\tau)$. In this pre-constraint regime, the coordinate time $t$ is a label, treated on the same footing as the spatial coordinates $\vec{x}$.

In the emergent Schr\"odinger representation, we consider a slice at fixed proper time $\tau$. The state is described by the extended functional
$\Psi[\phi,t,\tau]$, which represents the amplitude associated with the field configuration $\phi(\vec{x})$ at coordinate time $t$ and proper-time parameter $\tau$. Crucially, it is the cumulative action of the exact operator $\hat{F}^N$ on the coarse-grained macroscopic states that generates the effective unitary propagator in the IR limit. This asymptotically unitary continuous evolution can be matched with the structure of Ref.~\cite{Maiezza:2026wrp}, namely, the Nambu-Schr\"odinger functional equation for the extended states:
\begin{equation}
i \frac{\partial}{\partial \tau} \Psi[\phi,t,\tau] = H' \Psi[\phi,t,\tau] \,.
\end{equation}
This identification follows the same logic adopted in the cellular automaton framework of Ref.~\cite{thooft2016cellular}, where the quantum evolution operator is identified with the coarse-grained representation of the deterministic propagator.

Consistently, the continuous evolution in this extended space can be written as
\begin{equation}
\Psi[\phi,t,\tau]
=
\int \mathcal{D}\phi' dt' \,
K[\phi,t,\tau;\phi',t',0]\,
\Psi[\phi',t',0] \,.
\end{equation}
Finally, the physical post-constraint macroscopic description is obtained by saturating the proper-time evolution. As detailed in Ref.~\cite{Maiezza:2026wrp}, this is achieved by integrating the wavefunctional over $\tau$, with the lower limit set by the fundamental UV scale $\tau_{\min}$:
\begin{equation}
\Psi_R[\phi,t]
=
\int_{\tau_{\min}}^\infty d\tau \,
\Psi[\phi,t,\tau] \,.
\end{equation}
It is through this integration step that the extended pre-constraint description is projected onto the standard four-dimensional physical sector.

\subsection{Effective Planck Constant}\label{subsec:heff}

In what follows, we specialize, for simplicity, to the case in which the deterministic map $F$ has a single boundary-crossing microscopic configuration per elementary step,
\begin{equation}
\Delta n_\mu=1 \, ,
\end{equation}
rather than the generic case $\Delta n_\mu=\mathcal{O}(1)$. Thus, during one fundamental update, one microscopic path leaves the equivalence class and is mapped into the adjacent sector $[\phi_{\mu+1}]$. This situation is realized explicitly in the toy model of Appendix~\ref{appendix-toy}, where the macroscopic matrix elements are obtained directly from the microscopic deterministic kernel $\delta_{\phi',F(\phi)}$. The fraction of microscopic paths resolved as an outgoing boundary transition is therefore
\begin{equation}
\eta=\frac{1}{n_\mu}\,.
\end{equation}
When $n_\mu=1$, the coarse-grained class contains a single microscopic configuration. The system then undergoes a pure deterministic permutation, and Eq.~\eqref{K1} gives
\begin{equation}
\langle \phi_{\mu+1} \vert \hat F \vert \phi_\mu\rangle =1 \, .
\end{equation}
In this limit there are no unresolved internal alternatives inside the equivalence class, and the evolution is fully deterministic. Equivalently, no internal multiplicity is available to generate an effective quantum uncertainty at the coarse-grained level; see Appendix~\ref{appendix-toy} for a simplified realization.

For $n_\mu>1$, the macroscopic observer does not resolve the individual microscopic histories contained in $[\phi_\mu]$. The outgoing fraction $\eta=1/n_\mu$ is the part of the microscopic structure that is resolved as a boundary transition, while the complementary fraction
\begin{equation}
1-\eta = 1-\frac{1}{n_\mu}
\end{equation}
measures the amount of unresolved internal multiplicity inside the coarse-grained class. We identify this active unresolved fraction with the effective strength of quantum fluctuations. In this sense, the effective quantization scale is taken to be
\begin{equation}\label{heff_derived}
\hbar_{\rm eff}(\mu)
=
\hbar\left(1-\frac{1}{n_\mu}\right) \, .
\end{equation}
This expression should be understood as the minimal choice within the coarse-grained description adopted here. The only microscopic information
retained at this level is the resolved boundary fraction $\eta=1/n_\mu$, while the remaining fraction $1-\eta$ counts the
unresolved internal multiplicity of the equivalence class. Thus \eqref{heff_derived} is not introduced as a generic interpolation
between two limits, but as the direct normalization of the effective quantization strength by the unresolved sector. It gives
$\hbar_{\rm eff}=0$ for $n_\mu=1$, where the evolution is a purely deterministic permutation, and \(\hbar_{\rm eff}\to\hbar\) for $n_\mu\to\infty$, where the unresolved microscopic multiplicity becomes maximal.

More general non-minimal choices could be considered by replacing
$1-1/n_\mu$ with a function $Q(n_\mu)$ satisfying $Q(1)=0$ and $Q(n_\mu)\to1$ as $n_\mu\to\infty$. Such choices would amount to retaining additional microscopic information beyond the single boundary fraction \(\eta\), and would therefore define different
top-down completions. They may change the detailed behavior at intermediate scales, but the infrared recovery of the standard canonical structure is unchanged provided \(1-Q(n_\mu)\) vanishes sufficiently fast as $n_\mu\to\infty$. In the minimal counting model with $\Delta n_\mu=1$, the natural choice is precisely
$Q(n_\mu)=1-1/n_\mu$. This is different from situations in which a phenomenological ansatz is
truncated without a microscopic principle selecting the retained terms, a point emphasized for instance in discussions of GUP effective metrics and series-truncation ambiguities~\cite{Ong:2023jkp}.

The same expression has a natural entropic interpretation. Since
\begin{equation}
S_\mu=\ln n_\mu \, ,
\end{equation}
one has
\begin{equation}
1-\frac{1}{n_\mu}
=
1-e^{-S_\mu} \, .
\end{equation}
Therefore, $\hbar_{\rm eff}$ vanishes in the deterministic limit $S_\mu=0$ and approaches the standard Planck constant as the coarse-grained entropy grows.

The effective commutation relations are then normalized by this active quantum fraction. In the coarse-grained description we write
\begin{equation}\label{CCR_eff}
[\hat{\phi}(\vec{x}),\hat{\pi}(\vec{y})]
=
i\hbar_{\rm eff}(\mu)\delta(\vec{x}-\vec{y}) \, .
\end{equation}
For $n_\mu=1$, the effective variables describe a purely deterministic discrete evolution and the commutator vanishes. In the opposite limit, $n_\mu\to\infty$, the standard canonical commutation relations are recovered.
This differs from generalized-uncertainty-principle constructions, where the canonical commutator is usually deformed by momentum-dependent
corrections and the ultraviolet behavior depends on the chosen deformation function~\cite{Ong:2018zqn}. Here the modification is instead
a scalar coarse-graining factor fixed by the microscopic cardinality \(n_\mu\). The deformation of the canonical algebra is therefore not an
expansion in powers of momentum, but a consequence of projecting the deterministic dynamics onto unresolved equivalence classes.

Notice that, in a finite-dimensional Hilbert space, exact canonical commutation relations cannot hold because the trace of a commutator vanishes, as expressed by the Wielandt--Wintner obstruction. Equation~\eqref{CCR_eff} should therefore be understood as an effective continuum relation. The obstruction is removed in the macroscopic limit in which the number of equivalence classes becomes unbounded, corresponding to an unbounded proper-time domain $\tau\in[\tau_{\min},\infty)$. This is also the limit in which the effective generator can acquire a continuous spectrum.

This effective commutator may be understood as the coarse-grained remnant of the algebra of translations on the macroscopic equivalence class. In the simplified case $\Delta n_\mu=1$, one microscopic direction is resolved as a boundary transition, while the remaining $n_\mu-1$ directions are internal and indistinguishable at resolution $\mu$. The non-trivial canonical response is, therefore, weighted by the unresolved internal sector. Normalizing its dimension by the total cardinality of the class gives $ \frac{n_\mu-1}{n_\mu}=1-\frac{1}{n_\mu}$, which is precisely the factor multiplying $\hbar$ in Eq.~\eqref{heff_derived}. Thus Eq.~\eqref{CCR_eff} should be viewed as the macroscopic continuum limit of a projected coarse-grained algebra, rather than as a fundamental canonical relation at the deterministic level.

We now match this top-down expression to the running Planck constant obtained in the bottom-up formulation summarized in Sec.~\ref{sec:resume}. Comparing Eq.~\eqref{heff_derived} with Eq.~\eqref{hrun2}, and working in the local rest frame where $\tau=t$, one obtains the matching condition
\begin{equation}\label{matching_eq}
\frac{1}{n_\mu(\tau)}
=
4\,a\,\tau_{\min}\,\tilde f(\tau) \, ,
\qquad
\tau\geq\tau_{\min} \, .
\end{equation}
The right-hand side contains two a priori independent quantities: the normalization constant $a$, inherited from the freedom in normalizing the Nambu-like constraint, see Eq.~\eqref{gen3}, and the dimensionless proper-time kernel $\tilde f(\tau)$.

The matching at $\tau=\tau_{\min}$ should be understood as a boundary normalization at the first physical point of the proper-time domain, not as a perturbative statement about the validity of the effective expansion at the first microscopic step. At $\tau=\tau_{\min}$ no coarse-graining has yet occurred, and therefore
\begin{equation}
n_\mu(\tau_{\min})=1 \, .
\end{equation}
We choose the corresponding normalization of the kernel as
\begin{equation}
\tilde f(\tau_{\min})=1 \, .
\end{equation}
Evaluating Eq.~\eqref{matching_eq} at this boundary then fixes the otherwise arbitrary normalization constant,
\begin{equation}\label{fix_a}
4\,a\,\tau_{\min}=1 \, ,
\qquad\Rightarrow\qquad
a=\frac{1}{4\tau_{\min}} \, .
\end{equation}
With this normalization, the matching condition becomes
\begin{equation}
\tilde f(\tau)
=
\frac{1}{n_\mu(\tau)} \, .
\end{equation}
Thus the cutoff kernel of the bottom-up formulation is identified with the inverse cardinality of the coarse-grained equivalence class.

Using the power-law growth established in Sec.~\ref{sec:QFT-unit},
\begin{equation}
n_\mu
=
(1+\mu)^p
=
\left(\frac{\tau}{\tau_{\min}}\right)^p ,
\qquad
\mu=
\frac{\tau-\tau_{\min}}{\tau_{\min}}
\geq 0 \, ,
\end{equation}
with the normalization $n_\mu(\tau_{\min})=1$ built in, one obtains
\begin{equation}\label{true-behavior}
\tilde f(\tau)
=
\left(\frac{\tau_{\min}}{\tau}\right)^p ,
\qquad
\tau\geq\tau_{\min} \, .
\end{equation}
This expression satisfies $\tilde f(\tau_{\min})=1$ and $\tilde f(\tau)\to0$ as $\tau\to\infty$ for any $p>0$. The stronger condition $p>1$, derived in Sec.~\ref{sec:QFT-unit}, ensures that the cumulative dissipative correction vanishes in the infrared and that standard unitary QFT is recovered.

If a non-minimal choice $Q(n_\mu)$ were adopted, the matching to the bottom-up running Planck constant would be modified accordingly. For
example, if $1-Q(n_\mu)\sim n_\mu^{-\alpha}$ at large $n_\mu$, the proper-time kernel would scale as
\begin{equation}
\tilde f(\tau)\sim
\left(\frac{\tau_{\min}}{\tau}\right)^{p\alpha} \,.
\end{equation}
The minimal choice used in Eq.~\eqref{heff_derived} corresponds to $\alpha=1$. Thus the precise intermediate-scale profile is model-dependent, whereas the infrared conclusion is robust whenever $p\alpha>1$, so that the cumulative dissipative correction is washed out at large proper time.

It is useful to translate this result back into the $\lambda$-representation used in the Nambu-like formulation. With the Fourier convention
\begin{equation}
\tilde f(\tau)
=
\int_{-\infty}^{+\infty} d\lambda\,
f(\lambda)\,e^{i\lambda\tau} \, ,
\end{equation}
the power-law kernel in Eq.~\eqref{true-behavior}, restricted to the physical domain $\tau\geq\tau_{\min}$, defines the spectral profile of the off-constraint correction. Equivalently, one may associate to it the one-sided inverse transform
\begin{equation}
\chi_p(\lambda;\tau_{\min})
=
\frac{1}{2\pi}
\int_{\tau_{\min}}^\infty d\tau\,
\left(\frac{\tau_{\min}}{\tau}\right)^p
e^{-i\lambda\tau}
=
\frac{\tau_{\min}}{2\pi}
\int_1^\infty du\,u^{-p}\,
e^{-i\lambda\tau_{\min}u} \, .
\end{equation}
For $p>1$, this expression is well defined and depends on the dimensionless combination $\lambda\tau_{\min}$. It is concentrated around the constraint surface $\lambda=0$ and describes the off-constraint tail associated with the soft violation of the Nambu-like constraint.

Strictly speaking, this tail should not be identified with the full spectral weight of the physical state. The exact Nambu projection still contains the $\delta(\lambda)$ sector selected by the constraint. At finite $\tau_{\min}$ the spectral support is therefore schematically of the form
\begin{equation}
f_{\rm tot}(\lambda;\tau_{\min})
=
\delta(\lambda)
+
\chi_p(\lambda;\tau_{\min}) \, ,
\end{equation}
where $\chi_p$ denotes the off-constraint contribution whose Fourier transform gives the power-law kernel on the physical proper-time domain. In the limit in which the minimal proper-time deformation is removed, the off-constraint tail disappears in the distributional sense, and the standard $\delta(\lambda)$ projection of the Nambu construction is recovered.

Equation~\eqref{true-behavior} resolves, within the physical proper-time domain $\tau\geq\tau_{\min}$, an ambiguity left open in Ref.~\cite{Maiezza:2026wrp}, where $\tilde f$ was introduced as an external ansatz -- for instance a Gaussian, $\tilde f(t)=e^{-\sigma^2 t^2/2}$ -- chosen for mathematical tractability. Both forms play the same qualitative role: they are normalized at the onset of the ultraviolet regime and suppress the non-unitary correction toward the infrared. However, they differ structurally. The Gaussian ansatz decays exponentially in $t^2$, whereas the top-down result is an inverse power law controlled by the complexity exponent $p$. In particular, the free parameter $\sigma$ of the bottom-up ansatz is replaced by the exponent $p$, which is constrained by the macroscopic dimensionality requirement discussed in Sec.~\ref{sec:QFT-unit} and in the remarks below Eq.~\eqref{GR}. The cutoff kernel $\tilde f$ is thus promoted from an external input of the effective theory to a prediction of the microscopic deterministic dynamics.

\section{Toward emergent relativity in the macroscopic limit}\label{sec:GR}

The underlying Nambu-like dynamics allows, in the macroscopic infrared regime, an effective description compatible with smooth changes of coordinates,
$x'^\mu=x'^\mu(x)$. In this regime the coarse-grained cardinality becomes large, $n_\mu\to\infty$, and the dissipative contribution
$\Delta n_\mu/n_\mu$ is suppressed. The macroscopic dynamics is then effectively unitary and admits a diffeomorphism-covariant formulation. We now show how, within this infrared regime, the coarse-grained deterministic substrate can be associated with an effective smooth metric structure.

To formalize this point, we exploit the correspondence between emergent coordinate labels and fundamental events, schematically denoted by
\begin{equation}
x \leftrightarrow e_i^c \, .
\end{equation}
At this stage, $x$ should not be understood as a point of a pre-existing spacetime manifold. Rather, it is an effective coordinate label associated with the coarse-grained description of the microscopic event structure. The geometric interpretation arises only after the infrared coarse-graining has produced a smooth macroscopic regime.

The first step is to allow the cardinality of the coarse-grained equivalence classes to depend locally on the microscopic events,
\begin{equation}
n_\mu \longrightarrow n_\mu(x(e_i^c)) \, .
\end{equation}
This local promotion is not introduced as an external field on a pre-existing geometry. It expresses the fact that, in the coarse-grained description, different macroscopic regions may contain different numbers of microscopic deterministic histories that are unresolved by the observer. Thus $n_\mu(x)$ measures the local density of indistinguishable microscopic histories and plays the role of an internal geometric datum. In the infrared regime, where a smooth description becomes meaningful, this local density is the quantity from which the effective metric structure will be reconstructed.

We then define the coordinate fluctuation between two microscopic events as
\begin{equation}
\Delta x^\alpha
\equiv
x^\alpha(e_i^c)-x^\alpha(e_j^{c'}) \, .
\end{equation}
The explicit presence of distinct chain labels, $c$ and $c'$, is important. As anticipated in Sec.~\ref{sec:setup}, comparing events belonging to different chains is physically meaningful only because the microscopic dynamics $F$, together with the subsequent coarse-graining, interweaves the chains. If the chains were dynamically isolated, transversal fluctuations would be disconnected and no extended multi-dimensional macroscopic geometry could emerge.

Because of this interweaving, one can define an effective contravariant pre-metric as the statistical covariance of coordinate fluctuations between adjacent fundamental events, weighted by the local cardinality of unresolved microscopic histories:
\begin{equation}\label{gmunu}
g^{\alpha\beta}(x)
\equiv
\frac{1}{\tau_{\min}^2}
\langle \Delta x^\alpha \Delta x^\beta\rangle_{n_\mu}
\equiv
\frac{1}{\tau_{\min}^2}
\frac{
\sum_{\mathcal D}
\Delta x^\alpha\Delta x^\beta\,
n_\mu(x(e_j^{c'}))
}{
\sum_{\mathcal D}
n_\mu(x(e_j^{c'}))
}
\, .
\end{equation}
Here the summation domain $\mathcal D$ consists of all pairs $(j,c')$ such that the microscopic map connects $e_j^{c'}$ to $e_i^c$, or conversely $e_i^c$ to $e_j^{c'}$:
\begin{equation}
\mathcal D
=
\Bigl\{
(j,c')\; \Big|\;
F(e_j^{c'})=e_i^c
\;\;{\rm or}\;\;
F(e_i^c)=e_j^{c'}
\Bigr\} \, .
\end{equation}
This set is non-empty precisely because $F$ interweaves distinct chains, as specified in Eq.~\eqref{interweaving}. In the absence of such interweaving, the transverse covariance would vanish identically, and the construction would reduce to a collection of disconnected one-dimensional histories rather than an extended geometry.

The weighting by $n_\mu$ has a direct microscopic interpretation. In the coarse-grained description, two adjacent macroscopic events are not connected by a single microscopic link, but by a bundle of indistinguishable microscopic histories. The cardinality $n_\mu$ therefore acts as a measure of entropic connectivity, since
\begin{equation}
S_\mu=\ln n_\mu \, ,
\end{equation}
between neighboring coarse-grained events. The effective macroscopic geometry is consequently shaped by the local density of microscopic histories. In this sense, the metric is not introduced through an external gravitational action, but reconstructed internally from the statistical structure of the deterministic substrate.

Under a general coordinate transformation $x^\alpha\to x'^\alpha(x)$, the coordinate fluctuations transform, at leading order in the discrete separation, as
\begin{equation}
\Delta x'^\alpha
=
\frac{\partial x'^\alpha}{\partial x^\beta}
\Delta x^\beta
+
\mathcal O(\Delta x^2) \, .
\end{equation}
Consequently, in the macroscopic limit in which the discrete separations are small compared with the coarse-grained scale, the statistical matrix in Eq.~\eqref{gmunu} transforms as a rank-two contravariant tensor. When this covariance matrix is non-degenerate, its inverse defines the corresponding covariant object $g_{\alpha\beta}$. From this effective metric one can construct the associated connection, curvature tensors, and geometric quantities needed for the infrared gravitational description. This kinematic structure sets the stage for the use of Lovelock's theorem in the macroscopic limit~\cite{lovelock1971einstein,lovelock1972four}.

\paragraph{Euclidean versus Lorentzian signature.}

By construction, the covariance matrix in Eq.~\eqref{gmunu} is positive semi-definite and, in the non-degenerate case, defines a Riemannian pre-geometry. At this kinematic level all coordinate labels $x^\alpha=(x^0,\vec x)$ are treated on the same footing. In particular, $x^0$ is still only a coordinate label of the coarse-grained event structure, not yet the physical time variable of the post-constraint theory.

However, the fundamental substrate is not an unordered set of events. From the beginning, each microscopic chain is endowed with the ordering relation
\begin{equation}
e_i^c \prec e_j^c
\qquad
{\rm for}
\qquad
i<j \, ,
\end{equation}
as in Eq.~\eqref{event_order}. This ordering provides a discrete precursor of temporal orientation before the emergence of a continuous spacetime manifold. Therefore, the physical time variable is not introduced from nothing at the effective level: it is the macroscopic realization of an ordering already present in the microscopic substrate.

The physical Lorentzian signature is selected only after the Nambu-like constraint is imposed. The covariance construction provides the pre-geometric metric data, while the Nambu-like constraint selects the physical sector whose propagating modes obey a Lorentzian mass-shell condition. Thus, although Eq.~\eqref{gmunu} defines a Riemannian pre-geometry, the metric probed by on-shell excitations is Lorentzian. This transition can be represented as a dynamical Wick rotation of the coordinate associated with the microscopic ordering and with the monotonic growth of the coarse-graining scale $n_\mu$.

Indeed, the propagator of the post-constraint physical fields obtained in Ref.~\cite{Maiezza:2026wrp} obeys the standard mass-shell relation
\begin{equation}\label{mass_shell}
p_0^2-\vec p^{\,2}-m^2=0 \, ,
\end{equation}
which follows from the Nambu operator
\begin{equation}
H'=H-i\partial_t \, .
\end{equation}
This operator already distinguishes the label $t$ from the spatial coordinates $\vec x$. Therefore, the Lorentzian character of Eq.~\eqref{mass_shell} should not be regarded as an independent consequence of the Euclidean covariance matrix \eqref{gmunu} alone. Rather, it is anchored to the primitive ordering of the fundamental events, Eq.~\eqref{event_order}, and implemented at the macroscopic level by the Nambu-like constraint. The constraint acts as a selection rule that promotes the ordered microscopic label to the physical time variable of the post-constraint theory.

Within this interpretation, the mismatch between the Riemannian pre-geometry defined by the covariance matrix and the Lorentzian metric perceived by on-shell fields is represented by a Wick rotation of the ordered coordinate. The on-shell condition \eqref{mass_shell} motivates the identification
\begin{equation}
p_0^{\rm Eucl}=i p_0 \, ,
\end{equation}
or equivalently, in coordinate space,
\begin{equation}\label{Wick_rotation}
x^0_{\rm Eucl}
\longrightarrow
i t \, .
\end{equation}
Consequently,
\begin{equation}
(dx^0_{\rm Eucl})^2 \longrightarrow -dt^2 \, ,
\end{equation}
so that the pre-geometric line element
\begin{equation}
ds^2=(dx^0_{\rm Eucl})^2+d\vec x^{\,2}
\end{equation}
is mapped, in the physical on-shell sector, to the Lorentzian line element
\begin{equation}
ds^2=-dt^2+d\vec x^{\,2} \, .
\end{equation}
A fully self-contained derivation of the Lorentzian signature from the discrete network alone remains an open problem. In the present framework, the signature is instead selected by the combined structure of microscopic ordering and the Nambu-like physical constraint.

\paragraph{Einstein equations.}

In the infrared regime, $n_\mu\gg 1$, the dissipative corrections are suppressed and the coarse-grained evolution becomes effectively unitary. In this limit the Nambu-Schr\"odinger system reproduces the standard quantum-field-theoretic dynamics in the physical sector. Before the proper-time constraint is saturated, however, the effective description still lives in the extended configuration space $(\phi,t,\tau)$. Accordingly, the effective Nambu energy at fixed proper time $\tau$ is written as
\begin{equation}\label{Heff_exp}
\langle H_{\rm eff}\rangle_\tau
=
\int\!\mathcal{D}\phi\,dt\;
\Psi^*[\phi,t,\tau]\,
H\,
\Psi[\phi,t,\tau] \, .
\end{equation}
The explicit integration over the coordinate label $t$ reflects the pre-constraint nature of the description. The Hamiltonian $H$ becomes the physical generator of ordinary time evolution only after the proper-time integration has projected the system onto the physical sector.

Since the metric is not assumed as a fundamental background object, but reconstructed from the coarse-grained covariance of microscopic event fluctuations, the source tensor entering the macroscopic gravitational equations must also be defined internally within the emergent description. We therefore define the macroscopic stress-energy tensor as the linear response of the effective Nambu energy to a local deformation of the emergent metric data. Up to a conventional dimensionful normalization, we write
\begin{equation}\label{Tmunu}
T_{\alpha\beta}(x)
\equiv
\frac{\delta \langle H_{\rm eff}\rangle_\tau}
{\delta g^{\alpha\beta}(x)} \, .
\end{equation}
This definition should be understood as an effective response relation, rather than as a variation of a fundamental gravitational action. The quantity $\langle H_{\rm eff}\rangle_\tau$ in Eq.~\eqref{Heff_exp} is a functional of the emergent metric through the metric dependence of the infrared Hamiltonian, including its kinetic terms and integration measure. The overall normalization implicit in Eq.~\eqref{Tmunu} is not fixed by the present kinematical construction and will be absorbed into the coupling $\kappa$ appearing in the macroscopic gravitational equations.

In the same infrared regime, the suppression of the dissipative term restores the Ward identity associated with diffeomorphism covariance. Indeed, as $n_\mu\to\infty$, one has $\Delta n_\mu/n_\mu\to0$, and the coarse-grained dynamics becomes effectively unitary. Together with the diffeomorphism-covariant macroscopic description constructed above, this implies the covariant conservation law
\begin{equation}\label{conserveE}
\nabla^\alpha T_{\alpha\beta}=0 \, .
\end{equation}
This conservation equation is then matched, on the geometric side, to the contracted Bianchi identity.

We now ask which geometric tensor can balance the internally defined source $T_{\alpha\beta}$ in the infrared macroscopic limit. At this stage the role of Lovelock's theorem is not to derive the metric from the microscopic dynamics, but to select the admissible local gravitational equations once a smooth, non-degenerate, diffeomorphism-covariant metric has emerged. In four dimensions, assuming locality, second-order metric field equations, and no additional long-range tensorial structures beyond the emergent metric, Lovelock's theorem states that the only symmetric, divergence-free rank-two tensor constructed from the metric and its first and second derivatives is the Einstein tensor plus a cosmological constant term. Under these infrared assumptions, Eq.~\eqref{conserveE} and diffeomorphism covariance select the effective gravitational equations
\begin{equation}\label{GR}
G_{\alpha\beta}
+
\Lambda g_{\alpha\beta}
=
\kappa T_{\alpha\beta} \, .
\end{equation}
Here $\kappa=8\pi G_N$ is the macroscopic gravitational coupling. In the present framework it is not derived from a fundamental Einstein-Hilbert action, but left as a phenomenological parameter of the infrared theory, absorbing the normalization chosen in the definition of $T_{\alpha\beta}$. Appendix~\ref{appendix-gravity-toy} provides a toy model illustrating how local variations in the microscopic cardinality can induce a non-trivial effective geometry.

As $\tau\to\tau_{\min}$, equivalently $n_\mu\to1$, the statistical variance defining $g^{\alpha\beta}$ in Eq.~\eqref{gmunu} loses its continuum meaning, and the smooth geometric description breaks down. Classical gravitational singularities are therefore not resolved by extending the metric beyond its domain of validity. Rather, within this framework, the differentiable spacetime description ceases to apply before a curvature divergence can be physically defined. In this sense, the deterministic substrate provides a possible kinematical mechanism for avoiding the singular regime of classical general relativity.

\paragraph{Remarks}

\begin{itemize}
    
\item The geometric emergence discussed above provides a physical interpretation for the growth exponent $p$ and connects it to the convergence condition $p>1$ required for the recovery of emergent unitary QFT in Sec.~\ref{sec:QFT-unit}, around Eq.~\eqref{U-nonU}. In the underlying pre-geometric deterministic network, the causal horizon is expected to expand linearly with the number of dynamical steps $N$. Consequently, the number of accessible microstates within an indistinguishable macroscopic class scales as an effective volume, $n_\mu\sim N^d$, suggesting the identification $p\simeq d$.

    Rather than dynamically deriving the macroscopic dimensionality from first principles, we impose it as a phenomenological matching condition. If the target continuum limit is to describe our observable four-dimensional spacetime, augmented before the constraint by the proper-time parameter $\tau$, the relevant pre-constraint effective dimensionality is $d=5$. The underlying discrete network should then be characterized by $p=5$. This value automatically satisfies the convergence condition $p>1$, ensuring that the dissipative correction vanishes in the infrared and that the emergent unitary dynamics is self-consistent.

\item After the Nambu-like constraint is imposed, the four-dimensionality of the emergent physical manifold is a phenomenological input, not a result derived from first principles. Under this assumption, and within the standard hypotheses of Lovelock's theorem, General Relativity is selected as the effective macroscopic gravitational theory.

\end{itemize}

\paragraph{On QFT in curved spacetime.}

The cardinality of the equivalence class, once promoted to a spacetime-dependent field, implies that the matching condition in \eqref{matching_eq} becomes local:
\begin{equation}
\frac{1}{n_\mu(x)} \longleftrightarrow 4 a\,\tau_{\min}\,\tilde{f}(\tau(x)) \,.
\end{equation}
This local promotion has a direct implication for the effective commutation relations in Eq.~\eqref{CCR_eff}. The effective Planck constant appearing in the coarse-grained description becomes a local quantity,
$$
\hbar_{\rm eff}\longrightarrow \hbar_{\rm eff}(x)  \,,
$$
modulated by the local geometric structure through the dependence of $n_\mu$ on the accumulated proper time. In this sense, the local amount of ``quantumness'' -- and the corresponding residual loss of macroscopic unitarity -- is controlled by the spacetime dependence of the coarse-graining scale.

The spacetime dependence of the equivalence-class cardinality, $n_\mu\to n_\mu(x)$, also affects the generator of the effective macroscopic evolution. In the functional Schr\"odinger picture, the Hamiltonian density acquires an explicit dependence on the local geometric data, encoded here through the local coarse-graining structure:
\begin{equation}
i \frac{\partial \Psi[\phi,t]}{\partial t} = \int d^3x \, \mathcal{H}(x,\hat{\phi},\hat{\pi})\, \Psi[\phi,t] \,.
\end{equation}
In the standard continuum framework, such a spacetime-dependent Hamiltonian density, governed by the local flow of proper time and by the
background geometry, is one of the characteristic features of QFT in curved spacetime. Therefore, the promotion of the coarse-graining
parameter to a local field $n_\mu(x)$ does not only modify the effective commutation relations; it also induces the spacetime-dependent generator expected for a quantum field theory propagating on an emergent curved background.

\subsection{Medium-High energy regime corrections}

Equation~\eqref{GR} holds in the infrared regime, $n_\mu\gg 1$, where the dissipative correction induced by coarse-graining is suppressed and the macroscopic Ward identity is restored. At intermediate energies, however, the residual non-unitarity of the coarse-grained evolution produces a controlled violation of the conservation law \eqref{conserveE}.

In the simplified case considered in Sec.~\ref{subsec:heff}, one has $\Delta n_\mu=1$. The fractional leakage per elementary step is therefore
\begin{equation}
\frac{\Delta n_\mu(x)}{n_\mu(x)}
=
\frac{1}{n_\mu(x)} \, .
\end{equation}
Using the matching result \eqref{true-behavior}, this quantity is suppressed as
\begin{equation}
\frac{1}{n_\mu}
\sim
\left(\frac{\tau_{\min}}{\tau}\right)^p
\longrightarrow 0
\end{equation}
in the infrared limit. Thus any violation of the macroscopic conservation law must vanish as $n_\mu\to\infty$.

At leading order in the leakage fraction, we parametrize the corresponding non-conservation of the effective stress tensor by a source current controlled by the local variation of $1/n_\mu(x)$:
\begin{equation}\label{nonconserveE}
\nabla^\alpha T_{\alpha\beta}
=
\mathcal I_\beta \, ,
\qquad
\mathcal I_\beta
=
\zeta\,\nabla_\beta
\left(
\frac{1}{n_\mu(x)}
\right) \, .
\end{equation}
Here $\zeta$ is a dimensionful normalization constant fixed by the microscopic-to-macroscopic matching. In what follows it can be absorbed into the normalization of the effective correction, but keeping it explicit makes clear that Eq.~\eqref{nonconserveE} is a leading effective parametrization.

The form of $\mathcal I_\beta$ is motivated directly by the dissipative factor in the coarse-grained evolution. From Eq.~\eqref{Evolution}, the non-unitary contribution per elementary step is controlled by
\begin{equation}
\exp\left[-\frac{\Delta n_\mu}{n_\mu(x)}\right] \, .
\end{equation}
For $\Delta n_\mu=1$, the local strength of this correction is $1/n_\mu(x)$. Therefore, when the microscopic density of unresolved histories varies across the emergent geometry, the local leakage fraction also varies, and the Ward identity associated with the infrared diffeomorphism-covariant regime is violated by a term proportional to its gradient. At leading order one obtains
\begin{equation}
\nabla^\alpha T_{\alpha\beta}
=
\zeta\,\nabla_\beta
\left(
\frac{1}{n_\mu(x)}
\right)
+
\mathcal O(n_\mu^{-2}) \, .
\end{equation}
This expression satisfies the required infrared limit, since $\mathcal I_\beta\to0$ as $n_\mu\to\infty$, and the standard conservation law \eqref{conserveE} is recovered.

The modified macroscopic gravitational equation can then be written as
\begin{equation}\label{modGR}
G_{\alpha\beta}
+
\Lambda g_{\alpha\beta}
=
\kappa
\left(
T_{\alpha\beta}
+
\Theta_{\alpha\beta}
\right) \, ,
\end{equation}
where $\Theta_{\alpha\beta}$ is an effective compensating tensor whose role is to restore compatibility with the contracted Bianchi identity. Since
\begin{equation}
\nabla^\alpha
\left(
G_{\alpha\beta}
+
\Lambda g_{\alpha\beta}
\right)
=
0 \, ,
\end{equation}
the total source on the right-hand side of Eq.~\eqref{modGR} must be covariantly conserved. Hence $\Theta_{\alpha\beta}$ must satisfy
\begin{equation}
\nabla^\alpha \Theta_{\alpha\beta}
=
-
\mathcal I_\beta \, .
\end{equation}
Within the restricted class of corrections that introduce no additional long-range tensorial structures beyond the emergent metric, are isotropic in the local rest frame, and depend only on the scalar leakage variable $1/n_\mu(x)$, the minimal ansatz is
\begin{equation}
\Theta_{\alpha\beta}
=
f(x)\,g_{\alpha\beta} \, .
\end{equation}
Using metric compatibility, $\nabla_\gamma g_{\alpha\beta}=0$, one finds
\begin{equation}
\nabla^\alpha
\left(
f\,g_{\alpha\beta}
\right)
=
\nabla_\beta f \, .
\end{equation}
The compensating condition therefore reduces to
\begin{equation}
\nabla_\beta f
=
-
\zeta\,\nabla_\beta
\left(
\frac{1}{n_\mu(x)}
\right) \, .
\end{equation}
Up to an irrelevant integration constant, which can be absorbed into the cosmological term, this gives
\begin{equation}\label{Theta}
\Theta_{\alpha\beta}
=
-
\zeta\,
\frac{1}{n_\mu(x)}
g_{\alpha\beta} \, .
\end{equation}
Equivalently, by absorbing $\zeta$ into the normalization of the effective correction, one may write $\Theta_{\alpha\beta}= -n_\mu^{-1}(x)g_{\alpha\beta}$. This term vanishes in the infrared limit, $n_\mu\to\infty$, and Eq.~\eqref{GR} is recovered. Conversely, it becomes maximal near the fundamental scale, $n_\mu\simeq1$, where the smooth geometric description itself approaches the boundary of its validity.

Thus Eq.~\eqref{Theta} should be understood as the minimal isotropic compensating term associated with the leading leakage-induced violation of the matter Ward identity. More general corrections are possible if additional tensorial structures, anisotropies, or further microscopic variables are retained. The expression above is the simplest one compatible with the assumptions of the infrared emergent metric description.

Before concluding this section, a brief cosmological remark is in order. The local non-conservation of the macroscopic energy-momentum tensor, driven by spatial or temporal variations of the microstate density $n_\mu(x)$, provides a possible effective perspective on the early Universe. At extremely high energies, near the regime in which the classical cosmological singularity would appear, the system approaches the deterministic limit,
\begin{equation}
n_\mu\to1,
\qquad
\hbar_{\rm eff}\to0 \, .
\end{equation}
As the fundamental evolution proceeds, microscopic information becomes increasingly unresolved by a coarse-grained observer, and the cardinality of the equivalence classes grows.

In a Friedmann--Lemaitre--Robertson--Walker background, a rapidly varying leakage fraction,
\begin{equation}
\nabla_\alpha
\left(
\frac{1}{n_\mu}
\right)
\neq 0 \, ,
\end{equation}
can be interpreted as an effective exchange between the microscopic deterministic substrate and the macroscopic energy-momentum balance. Borrowing from the thermodynamics of spacetime and viscous cosmology~\cite{prigogine1989thermodynamics,murphy1973big}, such an entropy-producing effective contribution may be represented, at the macroscopic level, as a bulk-viscous component with negative effective pressure. This suggests a possible connection with a primordial accelerated phase, although a detailed cosmological implementation lies beyond the scope of the present work.

It is important, however, to distinguish this effective macroscopic interpretation from the more basic mechanism by which the singular regime is avoided in the present framework. The avoidance of the initial singularity is not primarily a consequence of the viscous description. Rather, as the deterministic substrate is approached and $n_\mu\to1$, the differentiable manifold described by $g^{\alpha\beta}$ loses its domain of validity and is replaced by the discrete algebraic network of fundamental events. Without a continuous spacetime geometry, a curvature singularity is no longer a well-defined object. In this sense, the framework suggests a kinematical mechanism by which the singular regime of classical general relativity is avoided.

\section{Outlook}\label{sec:limitations}

The construction developed in this work should be understood as an effective top-down realization of the minimal proper-time framework. Its aim is not to provide a complete microscopic theory of all quantum fields and gravity, but to identify a deterministic substrate whose coarse-grained dynamics can reproduce, in the infrared, the main structures of the minimal proper-time formulation. Several directions remain open for a more complete development of the framework.

A first direction concerns the microscopic class of deterministic maps. The recovery of infrared unitarity relies on the separation between a boundary-like leakage, $\Delta n_\mu=\mathcal{O}(1)$, and the bulk growth of the coarse-grained equivalence classes, $n_\mu(N)\sim N^p$. This behavior is explicitly realized in the toy model of Appendix~\ref{appendix-toy}. It would be interesting to characterize more explicitly the locality properties of the pair $(F,C_\mu)$ under which the boundary leakage remains bounded, $\Delta n_\mu=\mathcal O(1)$, while the bulk cardinality of the coarse-grained classes grows as $n_\mu(N)\sim N^p$ under the repeated action of the fixed microscopic map.

A second direction concerns the effective operator algebra. The same counting argument that leads to the effective Planck constant also motivates the normalization of the macroscopic canonical commutator. In the present construction, the active quantum fraction is identified with the unresolved part of the coarse-grained equivalence class,
$$
\hbar_{\rm eff}(\mu)
=
\hbar\left(1-\frac{1}{n_\mu}\right).
$$
In the simplified case $\Delta n_\mu=1$, one microscopic direction is resolved as a boundary transition, while the remaining $n_\mu-1$ directions are internal and indistinguishable at resolution $\mu$. Therefore, the effective canonical response associated with coarse-grained translations is weighted by the same unresolved fraction, $\frac{n_\mu-1}{n_\mu} = 1-\frac{1}{n_\mu}$. Thus the macroscopic CCR can be interpreted as the continuum limit of a coarse-grained translation algebra normalized by the unresolved internal multiplicity of the equivalence class. Making this projected translation algebra fully explicit would provide a constructive realization of the emergent canonical structure.

A future challenge for the present framework is the explicit reproduction of non-local quantum correlations. Since the microscopic evolution governed by $F$ is deterministic and bijective, the theory bypasses the standard conclusions of Bell's theorem through the violation of Measurement Independence \cite{thooft2016cellular,Hossenfelder:2019shy}. In this context, such a violation is not a fine-tuning requirement but a structural consequence of the "interweaving" mechanism described in Sec. 3.1. Because the map $F$ connects microstates belonging to distinct causal chains (see Eq. \ref{interweaving}), the state of a particle and the setting of a macroscopic detector are both emergent features of the same underlying deterministic network. 

Consequently, the observed Bell-type correlations should be understood as topological constraints imposed by the global consistency of the microscopic map $F$ on the coarse-grained equivalence classes. Reproducing the precise quantitative form of these correlations (e.g., the Tsirelson bound) likely requires additional structure in the deterministic map, such as high non-factorizability between the variables associated with different emergent field species (see Sec. \ref{sec:setup}). Characterizing the specific classes of permutations that generate the correct infrared correlations remains a primary objective for the completion of the top-down program.

A third direction concerns the geometric sector. The covariance construction of Sec.~\ref{sec:GR} provides a Riemannian pre-geometric object, while the physical Lorentzian signature is selected by the microscopic ordering of events together with the Nambu-like constraint. A more complete analysis should clarify how this Lorentzian structure emerges from the discrete network in a fully intrinsic way. Similarly, while Lovelock's theorem selects Einstein gravity once a smooth, non-degenerate, diffeomorphism-covariant metric has emerged, the microscopic origin of the macroscopic parameters $\kappa$, $\Lambda$, and possible higher-curvature corrections remains to be understood.

Finally, the extension to realistic quantum field theories requires a richer microscopic substrate. One should allow for deterministic variables corresponding, in the infrared, to scalar, gauge, and fermionic degrees of freedom, for instance
\begin{equation}
\Phi_{\rm tot}(e_n)=\left(\phi^I(e_n),A_\mu^a(e_n),\psi^\alpha(e_n),\dots\right).
\end{equation}
In such a setting, gauge symmetries, fermionic statistics, and interactions should arise as emergent constraints on the coarse-grained dynamics, rather than as microscopic assumptions. This likely requires highly non-factorizable deterministic maps mixing the microscopic variables associated with different emergent field species.

These open directions do not affect the main purpose of the present work: to show that the minimal proper-time deformation, the running effective Planck constant, and an infrared relativistic geometric description can be given a common interpretation in terms of coarse-graining over deterministic microscopic histories.

\section{Summary}\label{sec:end}

In this work, we have proposed a top-down microscopic counterpart of the minimal proper-time formulation of quantum field theory. Starting from a deterministic substrate of causally ordered events, we have shown how coarse-graining over microscopic histories can lead to an effective Nambu-like dynamics at macroscopic scales. In this picture, the fundamental evolution remains deterministic and information-preserving, while the apparent loss of unitarity arises only because a coarse-grained observer cannot resolve all microscopic configurations.

The central mechanism is the growth of the coarse-grained equivalence classes. As the number of unresolved microscopic histories increases, the dissipative contribution induced by boundary leakage becomes progressively suppressed. In the infrared regime, this suppression allows the recovery of standard unitary QFT. Conversely, near the fundamental proper-time scale, the equivalence classes shrink toward the deterministic limit, and the effective quantum description ceases to be the appropriate one.

This same coarse-grained structure also provides a microscopic interpretation of the running effective Planck constant and of the proper-time cutoff kernel introduced in the bottom-up formulation of Ref.~\cite{Maiezza:2026wrp}. The cutoff is no longer an external ansatz, but is related to the inverse growth of unresolved deterministic histories. In this sense, the minimal proper-time deformation acquires a concrete statistical meaning: it measures the residual imprint of microscopic distinguishability on the macroscopic quantum dynamics.

We have also argued that the deterministic substrate can support an effective relativistic geometric description in the infrared. The metric is not introduced as a fundamental background object, nor derived from an external Einstein-Hilbert action. Instead, it is reconstructed from the weighted covariance of coordinate fluctuations between interwoven microscopic event chains. Once a smooth, non-degenerate, diffeomorphism-covariant metric description has emerged, Einstein gravity is selected as the effective macroscopic theory under the standard infrared assumptions summarized by Lovelock's theorem.

The framework therefore suggests a unified interpretation of two familiar ultraviolet problems. The divergences of continuum QFT and the singularities of classical general relativity both arise when macroscopic effective descriptions are extrapolated beyond their domain of validity. In the present picture, the continuum quantum and geometric descriptions break down near the deterministic microscopic regime, where the appropriate degrees of freedom are no longer fields on a smooth spacetime, but ordered events and their coarse-grained relations.

Overall, the construction points toward a common deterministic underpinning of minimal-scale structure, quantum behavior, and relativistic spacetime. Quantum field theory and general relativity then appear as complementary infrared descriptions of a deeper microscopic dynamics, accessed only after coarse-graining over fundamental events.

%%%%%%%%%%%%%%%%%%%%%%%%%%%%%%%%%%%%%%%%%%%%%%%%%%%%%%%%%%%%%%%%%%%%%%%%%%%%%%%%%%%%%%%%%%%%%%%%%%%%%%%%%%%%%%%%%%%%%%%%%%%%%%%%%%%%%

\section*{Acknowledgments}

I thank Juan Carlos Vasquez for his comments in the early stages of this work. The research is partially financed by INFN.

\appendix

\section{Simplest Toy Model}\label{appendix-toy}

Let us consider a simplified model in which the fundamental substrate consists of an ordered set of $L$ microstates, partitioned into three
macroscopic equivalence classes (or sectors), $\phi_\mu\in\{A,B,C\}$. For simplicity, we assume that each class contains the same number $n$ of microstates, so that $L=3n$.

\paragraph{Microscopic Map and Coarse-Grained Transition.}

The fundamental map $F$ is a cyclic permutation of the microscopic states,
\begin{equation}
F(\phi_k)=\phi_{k+1}\qquad {\rm mod}\,L \, .
\end{equation}
The coarse-grained transition amplitudes, acting on the effective macroscopic Hilbert space
$\mathrm{span}\{|A\rangle,|B\rangle,|C\rangle\}$, are constructed by evaluating the exact microscopic operator $\hat F$ between the
coarse-grained linear representations of the macroscopic equivalence classes. For a single step $\tau_{\min}$, the transition matrix elements
are
\begin{equation}
K_{ij}
=
\langle \phi_{\mu,i}|\hat F|\phi_{\mu,j}\rangle
=
\frac{1}{n}
\sum_{\phi\in[\phi_{\mu,j}]}
\sum_{\phi'\in[\phi_{\mu,i}]}
\delta_{\phi',F(\phi)} \, .
\end{equation}
Defining the scale parameter $\eta=1/n$, the matrix representation of the
macroscopic transition becomes
\begin{equation}
K
=
\begin{pmatrix}
1-\eta & 0 & \eta \\
\eta & 1-\eta & 0 \\
0 & \eta & 1-\eta
\end{pmatrix}
=
\mathbb{I}-\eta M\,,
\qquad
M
=
\begin{pmatrix}
1 & 0 & -1 \\
-1 & 1 & 0 \\
0 & -1 & 1
\end{pmatrix}.
\end{equation}
This explicitly illustrates the boundary nature of the leakage: during one elementary step, only one microscopic path leaves a given equivalence class and enters the adjacent one, while the remaining $n-1$ paths stay inside the same macroscopic sector. Thus this toy model realizes the unit-normalized boundary-leakage case,
\begin{equation}
\Delta n_\mu=1,
\qquad
\eta=\frac{\Delta n_\mu}{n_\mu}=\frac{1}{n}\, .
\end{equation}

\paragraph{Decomposition into Unitary and Dissipative Generators.}

To connect this explicit transition matrix with the parameterized macroscopic evolution used in the main text, we consider one fundamental step $\tau_{\min}$. The comparison is meaningful in the regime in which the deviation from the identity is small. In the present model this is controlled by $\eta=1/n$, and is therefore naturally realized for a large number of microscopic states in each equivalence class.

Expanding the parameterized macroscopic amplitude for small $\tau_{\min}H'$ and small dissipative correction, one obtains
\begin{equation}
K
\approx
\exp[-i\tau_{\min}H'-\mathcal D]
\approx
\mathbb{I}-i\tau_{\min}H'-\mathcal D \, ,
\end{equation}
where $\mathcal D$ denotes the dissipative generator induced on the coarse-grained macroscopic subspace. Its overall scale is controlled by the entropy variation
\begin{equation}
\Delta S_\mu
\simeq
\frac{\Delta n_\mu}{n_\mu}
=
\eta \, .
\end{equation}
On the other hand, the explicit transition matrix is $K=\mathbb{I}-\eta M$. Decomposing $M$ into its symmetric and antisymmetric parts,
$M=M_{\rm sym}+M_{\rm anti}$, gives
\begin{align}
M_{\rm sym}
&=
\frac{1}{2}
\begin{pmatrix}
2 & -1 & -1 \\
-1 & 2 & -1 \\
-1 & -1 & 2
\end{pmatrix},
\label{SymMatrix}
\\
M_{\rm anti}
&=
\frac{1}{2}
\begin{pmatrix}
0 & 1 & -1 \\
-1 & 0 & 1 \\
1 & -1 & 0
\end{pmatrix}.
\label{AntiMatrix}
\end{align}
Comparing the two expressions,
\begin{equation}
\mathbb{I}-i\tau_{\min}H'-\mathcal D
=
\mathbb{I}-\eta M_{\rm anti}-\eta M_{\rm sym}\,,
\end{equation}
one obtains the identifications
\begin{itemize}
    \item \textbf{Unitary sector:}
    \[
    i\tau_{\min}H'=\eta M_{\rm anti}\,,
    \qquad
    H'=-\frac{i\eta}{\tau_{\min}}M_{\rm anti}\,.
    \]
    Since $M_{\rm anti}$ is real and antisymmetric, the operator
    $-iM_{\rm anti}$ is Hermitian. This part therefore generates the
    phase rotation of the effective macroscopic dynamics.

    \item \textbf{Dissipative sector:}
    \[
    \mathcal D=\eta M_{\rm sym}\,.
    \]
    The dissipative generator is controlled by the same parameter
    $\eta=1/n$. In particular, the overall scale of the relative loss of microscopic
    distinguishability per step is $\Delta n/n=1/n$, matching the
    statistical estimate $\Delta S\simeq\Delta n/n$ for $\Delta n=1$.
\end{itemize}
Thus, in this toy model, the single parameter $\eta=1/n$ controls both the phase-generating and the dissipative parts of the coarse-grained macroscopic evolution. In the notation of Sec.~\ref{sec:QFT-unit}, this corresponds to the unit-normalized boundary-leakage case, with $\Delta n_\mu=1$ and an $\mathcal{O}(1)$ coefficient in the cumulative dissipative scaling. As $n\to\infty$, the macroscopic transition matrix approaches the identity for one elementary step, allowing a continuous differential evolution to emerge.

The spectrum of the Hermitian operator $-iM_{\rm anti}$ is
\[
-\frac{\sqrt{3}}{2},\quad 0,\quad \frac{\sqrt{3}}{2}\,.
\]
Consequently, in this toy model the corresponding eigenvalues of the Nambu-Schr\"odinger generator $H'$ are
\[
\lambda_k =
\frac{\eta}{\tau_{\min}} \left\{
-\frac{\sqrt{3}}{2},\, 0,\, \frac{\sqrt{3}}{2} \right\}.
\]
These eigenvalues should be interpreted as pre-constraint $\lambda$-eigenvalues of the extended Nambu-Schr\"odinger operator, rather than as ordinary energy levels. In particular, the $\lambda=0$ sector has an invariant meaning: it is the sector selected by the exact Nambu-like constraint. When the minimal proper time $\tau_{\min}$ is introduced, this constraint is softly relaxed, so that $\lambda\neq0$ sectors may contribute as controlled, UV-suppressed corrections. Therefore, unlike in the standard cogwheel discussion, no zero-point energy shift is required as part of the physical construction. The role of the cyclic deterministic map is instead to show explicitly how phase-generating eigenvalues of $H'$ can arise from an underlying permutation dynamics, leading to complex phase factors in the coarse-grained effective evolution.

Finally, choosing as an example $n=5$, the array $\delta_{\phi',F(\phi)}$ in Eq.~\eqref{K1} reads
\[
\left(
\begin{array}{ccc}
 \left(
\begin{array}{ccccc}
 0 & 0 & 0 & 0 & 0 \\
 1 & 0 & 0 & 0 & 0 \\
 0 & 1 & 0 & 0 & 0 \\
 0 & 0 & 1 & 0 & 0 \\
 0 & 0 & 0 & 1 & 0 \\
\end{array}
\right)
&
\left(
\begin{array}{ccccc}
 0 & 0 & 0 & 0 & 0 \\
 0 & 0 & 0 & 0 & 0 \\
 0 & 0 & 0 & 0 & 0 \\
 0 & 0 & 0 & 0 & 0 \\
 0 & 0 & 0 & 0 & 0 \\
\end{array}
\right)
&
\left(
\begin{array}{ccccc}
 0 & 0 & 0 & 0 & 1 \\
 0 & 0 & 0 & 0 & 0 \\
 0 & 0 & 0 & 0 & 0 \\
 0 & 0 & 0 & 0 & 0 \\
 0 & 0 & 0 & 0 & 0 \\
\end{array}
\right)
\\
 \left(
\begin{array}{ccccc}
 0 & 0 & 0 & 0 & 1 \\
 0 & 0 & 0 & 0 & 0 \\
 0 & 0 & 0 & 0 & 0 \\
 0 & 0 & 0 & 0 & 0 \\
 0 & 0 & 0 & 0 & 0 \\
\end{array}
\right)
&
\left(
\begin{array}{ccccc}
 0 & 0 & 0 & 0 & 0 \\
 1 & 0 & 0 & 0 & 0 \\
 0 & 1 & 0 & 0 & 0 \\
 0 & 0 & 1 & 0 & 0 \\
 0 & 0 & 0 & 1 & 0 \\
\end{array}
\right)
&
\left(
\begin{array}{ccccc}
 0 & 0 & 0 & 0 & 0 \\
 0 & 0 & 0 & 0 & 0 \\
 0 & 0 & 0 & 0 & 0 \\
 0 & 0 & 0 & 0 & 0 \\
 0 & 0 & 0 & 0 & 0 \\
\end{array}
\right)
\\
 \left(
\begin{array}{ccccc}
 0 & 0 & 0 & 0 & 0 \\
 0 & 0 & 0 & 0 & 0 \\
 0 & 0 & 0 & 0 & 0 \\
 0 & 0 & 0 & 0 & 0 \\
 0 & 0 & 0 & 0 & 0 \\
\end{array}
\right)
&
\left(
\begin{array}{ccccc}
 0 & 0 & 0 & 0 & 1 \\
 0 & 0 & 0 & 0 & 0 \\
 0 & 0 & 0 & 0 & 0 \\
 0 & 0 & 0 & 0 & 0 \\
 0 & 0 & 0 & 0 & 0 \\
\end{array}
\right)
&
\left(
\begin{array}{ccccc}
 0 & 0 & 0 & 0 & 0 \\
 1 & 0 & 0 & 0 & 0 \\
 0 & 1 & 0 & 0 & 0 \\
 0 & 0 & 1 & 0 & 0 \\
 0 & 0 & 0 & 1 & 0 \\
\end{array}
\right)
\end{array}
\right) .
\]
This explicit array shows that this toy model realizes the case in which only one microscopic path leaves a given equivalence class during a single elementary step. In the block representation above, this appears as a single non-vanishing entry in the adjacent off-diagonal block, while the remaining transitions stay inside the diagonal block. This is the toy-model realization of $\Delta n_\mu=1$.

\section{Toy Model for Effective Geometry}\label{appendix-gravity-toy}

We now present a minimal toy model illustrating how a non-trivial effective geometry can arise from a local dependence of the coarse-grained cardinality. In contrast with the QFT toy model of Appendix~\ref{appendix-toy}, where the cardinality was taken to be uniform inside each macroscopic sector, we now promote the same quantity to a local function of the emergent coordinate,
\begin{equation}
n_\mu \longrightarrow n_\mu(x) \, .
\end{equation}
We work in the infrared regime, where
\begin{equation}
\frac{\Delta n_\mu}{n_\mu}\longrightarrow 0 \, ,
\end{equation}
so that the dissipative contribution is suppressed and the macroscopic description is effectively unitary.

\paragraph{Discrete configuration and local cardinalities.}

We consider three coarse-grained sites along one spatial direction,
\begin{equation}
x_1=a\,,
\qquad
x_2=2a\,,
\qquad
x_3=3a\,,
\end{equation}
while the other coordinate labels are kept fixed,
\begin{equation}
t_i=y_i=z_i=0\,.
\end{equation}
Thus
\begin{equation}
X_i^\alpha=(0,x_i,0,0)\,.
\end{equation}
The local cardinalities are denoted by
\begin{equation}
n_i\equiv n_\mu(x_i)\,.
\end{equation}
The essential difference with respect to the QFT toy model is that the $n_i$ are now local data. A non-trivial dependence of $n_\mu(x)$ represents the microscopic input from which a non-trivial macroscopic geometry will be reconstructed.

\paragraph{Metric from the covariance.}

The effective contravariant metric data are constructed from the weighted covariance of coordinate fluctuations,
\begin{equation}
g^{\alpha\beta}(x_i)
=
\frac{1}{\tau_{\min}^2}
\frac{
\sum_{j\in D_i}
n_j\,\Delta X_{ij}^{\alpha}\Delta X_{ij}^{\beta}
}{
\sum_{j\in D_i}n_j
}\,,
\qquad
\Delta X_{ij}^{\alpha}=X_i^\alpha-X_j^\alpha\,.
\label{ToyMetricCovariance}
\end{equation}
At this stage Eq.~\eqref{ToyMetricCovariance} should be understood as the covariance prescription for the emergent metric data. The interpretation in terms of a smooth metric becomes meaningful only after passing to the infrared continuum approximation.

Since the present configuration probes only the $x$ direction, the non-trivial component fixed directly by Eq.~\eqref{ToyMetricCovariance} is
\begin{equation}
g^{xx}(x_i)
=
\frac{1}{\tau_{\min}^2}
\frac{
\sum_{j\in D_i}
n_j\,(x_i-x_j)^2
}{
\sum_{j\in D_i}n_j
}\,.
\end{equation}
For the three-site cyclic neighbourhood one obtains
\begin{align}
g^{xx}(a)
&=
\frac{a^2}{\tau_{\min}^2}
\frac{n_2+4n_3}{n_2+n_3}\,,
\\
g^{xx}(2a)
&=
\frac{a^2}{\tau_{\min}^2}\,,
\\
g^{xx}(3a)
&=
\frac{a^2}{\tau_{\min}^2}
\frac{4n_1+n_2}{n_1+n_2}\,.
\end{align}
We define
\begin{equation}
q_i\equiv g^{xx}(x_i)\,.
\end{equation}
In the continuum approximation the three values $q_i$ are regarded as samples of a smooth function,
\begin{equation}
q(x_i)=q_i\,.
\end{equation}
Therefore the covariance construction gives
\begin{equation}
g^{xx}(x)=q(x)\,,
\qquad
g_{xx}(x)=\frac{1}{q(x)}\,.
\end{equation}
Thus the function $q(x)$ is fixed by the local microscopic cardinalities through the covariance prescription.

To obtain a non-degenerate spacetime metric one must complete the covariance data. As discussed in the main text, the covariance matrix associated with microscopic fluctuations is naturally a Riemannian pre-geometric object. The physical Lorentzian signature is selected only after the Nambu-like constraint has picked the on-shell sector. Equivalently, the coordinate associated with the microscopic ordering is Wick-rotated to the physical time direction. In the present toy model this amounts to a Lorentzian completion of the covariance data, rather than to an independent derivation of the signature from the finite three-site system itself.

The configuration singles out the $x$ direction, but does not distinguish $y$ from $z$; we therefore impose residual isotropy in the transverse plane. We also work in the local diagonal frame, so that mixed components are absent. The static Lorentzian completion of the metric is then chosen as
\begin{equation}
g_{\mu\nu}(x)
=
{\rm diag}
\left(
-q(x),\,\frac{1}{q(x)},\,1,\,1
\right),
\label{ToyGRMetric}
\end{equation}
or, equivalently,
\begin{equation}
g^{\mu\nu}(x)
=
{\rm diag}
\left(
-\frac{1}{q(x)},\,q(x),\,1,\,1
\right).
\end{equation}
The relative minus sign in the time component encodes the Lorentzian projection selected by the post-constraint physical sector. The equality of the $y$ and $z$ components follows from the imposed residual transverse isotropy, and the absence of off-diagonal components follows from the choice of the local diagonal frame.

The corresponding line element is
\begin{equation}
ds^2
=
-q(x)\,dt^2
+
\frac{dx^2}{q(x)}
+
dy^2+dz^2\,.
\label{ToyGRLineElement}
\end{equation}
Thus the same function $q(x)$, determined by the local microscopic cardinalities through Eq.~\eqref{ToyMetricCovariance}, controls the macroscopic geometry. The flat case is recovered when $q(x)$ is constant, up to a trivial rescaling of the coordinates.

\paragraph{Curvature.}

For the metric in Eq.~\eqref{ToyGRMetric}, the non-vanishing Christoffel symbols are
\begin{equation}
\Gamma^{t}_{tx}
=
\Gamma^{t}_{xt}
=
\frac{q'}{2q}\,,
\qquad
\Gamma^{x}_{tt}
=
\frac{1}{2}qq'\,,
\qquad
\Gamma^{x}_{xx}
=
-\frac{q'}{2q}\,,
\end{equation}
where the prime denotes differentiation with respect to $x$. The non-vanishing components of the Ricci tensor are
\begin{equation}
R_{tt}
=
\frac{1}{2}qq''\,,
\qquad
R_{xx}
=
-\frac{1}{2}\frac{q''}{q}\,.
\end{equation}
The Ricci scalar is
\begin{equation}
R=-q''\,.
\end{equation}
Therefore,
\begin{equation}
G_{tt}=0\,,
\qquad
G_{xx}=0\,,
\qquad
G_{yy}=G_{zz}=\frac{1}{2}q''\,.
\end{equation}
In mixed components,
\begin{equation}\label{ToyEinsteinTensor}
G^\mu{}_\nu
=
{\rm diag}
\left(
0,\,
0,\,
\frac{1}{2}q'' ,\,
\frac{1}{2}q''
\right).
\end{equation}
At the central point one may estimate the second derivative by the finite difference
\begin{equation}\label{discreteD}
q''(2a) \simeq \frac{q(a)-2q(2a)+q(3a)}{a^2}  \,.
\end{equation}
This explicitly shows how local variations of the microscopic cardinalities, through their effect on $q(x)$, generate curvature in the infrared description.

\paragraph{Effective source and Einstein equations.}

In the infrared regime the effective stress tensor is covariantly conserved,
\begin{equation}
\nabla_\mu T^\mu{}_\nu=0\,.
\end{equation}
The effective source is not introduced as an independent matter sector on a pre-existing spacetime. Rather, consistently with the main text, it is defined internally as the response of the effective Nambu energy to local deformations of the emergent metric data.

For the metric in Eq.~\eqref{ToyGRMetric} one has
\begin{equation}
N=\sqrt{q(x)}\,,
\qquad
h_{ij}
=
{\rm diag}
\left(
\frac{1}{q(x)},\,1,\,1
\right),
\qquad
\sqrt{h}=\frac{1}{\sqrt{q(x)}}\,.
\end{equation}
Thus, for an emergent scalar field, the infrared Hamiltonian entering the expectation value can be written as
\begin{equation}
H_\phi[q]
=
\int d^3x
\left[
\frac{q(x)}{2}\pi^2
+
\frac{q(x)}{2}(\partial_x\phi)^2
+
\frac{1}{2}(\partial_y\phi)^2
+
\frac{1}{2}(\partial_z\phi)^2
+
V(\phi)
\right].
\label{ToyScalarHamiltonian}
\end{equation}
The corresponding Nambu-Schr\"odinger expectation value is
\begin{equation}
\langle H_{\rm eff}[q]\rangle_\tau
=
\int D\phi\,dt\,
\Psi^*[\phi,t,\tau]\,
H_\phi[q]\,
\Psi[\phi,t,\tau]\,.
\label{ToyHeffExpectation}
\end{equation}
The macroscopic stress tensor is then defined by the linear response
\begin{equation}
T_{\mu\nu}(x)
\equiv
\frac{\delta \langle H_{\rm eff}[q]\rangle_\tau}
{\delta g^{\mu\nu}(x)}\,,
\label{ToyStressResponse}
\end{equation}
up to the conventional normalization absorbed into $\kappa$. This should be understood as an effective response relation internal to the emergent description, not as the variation of a fundamental matter action on a background metric.

Since the background is static, diagonal, and invariant under $y\leftrightarrow z$, this response has the mixed form
\begin{equation}
T^\mu{}_\nu
=
{\rm diag}
\left(
-\rho,\,
p_x,\,
p_\perp,\,
p_\perp
\right).
\label{ToyStressTensor}
\end{equation}
The functions $\rho$, $p_x$ and $p_\perp$ therefore denote the independent components of the response of $\langle H_{\rm eff}[q]\rangle_\tau$ to metric deformations in this static, diagonal, transversely isotropic background.

The infrared Einstein equations are
\begin{equation}
G^\mu{}_\nu+\Lambda\delta^\mu{}_\nu
=
\kappa T^\mu{}_\nu\,.
\end{equation}
Using Eq.~\eqref{ToyEinsteinTensor}, one obtains
\begin{align}
\Lambda
&=
-\kappa\rho\,,
\\
\Lambda
&=
\kappa p_x\,,
\\
\frac{1}{2}q''+\Lambda
&=
\kappa p_\perp\,.
\end{align}
For the illustrative choice $\Lambda=0$, this gives
\begin{equation}
\rho=0\,,
\qquad
p_x=0\,,
\qquad
p_\perp
=
\frac{1}{2\kappa}q''\,,
\end{equation}
where, from \eqref{discreteD}, one has for $q''$ the discrete estimation
\begin{equation}
q'' \to \frac{3 \left(n_2 n_3+n_1 \left(n_2+2 n_3\right)\right)}{\left(n_1+n_2\right) \left(n_2+n_3\right) \tau^2}  \,.
\end{equation}
Thus, a local variation of the microscopic cardinality $n_\mu(x)$ induces a position-dependent metric function $q(x)$, whose curvature is supported by an effective transverse pressure in the infrared response description. When $q''=0$, and in particular when $q$ is constant, the flat infrared geometry is recovered.
\vspace{2em}

\bibliography{biblio}% common bib file
%% if required, the content of .bbl file can be included here once bbl is generated
%%\input sn-article.bbl

\end{document}